\documentclass{article}
\usepackage[numbers,sort&compress]{natbib}
\usepackage{todonotes}
\usepackage{geometry}
\usepackage{amsmath}

\usepackage{hyperref}
\usepackage[title]{appendix}

\numberwithin{figure}{section}
\numberwithin{table}{section}
\numberwithin{equation}{subsection}
\usepackage{float} 
\usepackage{amssymb}
\usepackage{amsthm}
\usepackage{pifont}
\usepackage{bm}
\usepackage{graphicx}
\graphicspath{{./figures/}}
\usepackage{color}
\usepackage{tabu}
\usepackage{booktabs}
\usepackage{caption}
\usepackage{authblk}
\usepackage{subfigure}
\usepackage{enumerate}
\usepackage{tabu}
\usepackage{multirow}
\usepackage{diagbox}
\usepackage{tikz}
\usepackage{algorithm,algorithmic}

\begin{document}
\title{RT-APNN for Solving Gray Radiative Transfer Equations
\footnotetext{
\endgraf{{\scriptsize $^1$} Institute of Applied Physics and Computational Mathematics,
Beijing 100088, China.}
\endgraf{{\scriptsize $^2$} National Key Laboratory of Computational Physics, Beijing 100088, China.}
\endgraf{{\scriptsize $^3$} School of Mathematical Sciences, Shanghai Jiao Tong University, Shanghai 200240, China.}
\endgraf{{\scriptsize $^4$} HEDPS, Center for Applied Physics and Technology, College of Engineering, Peking University, Beijing 100871, China.}

\endgraf{{\scriptsize $^*$} Corresponding author.}
\endgraf{E-mail address: xiexizhe21@gscaep.ac.cn, chenwg@iapcm.ac.cn, zhengma@sjtu.edu.cn, wang\_han@iapcm.ac.cn.}
\endgraf{The work of W. Chen is supported partly by the NSFC No. 12271050, 
Foundation of National Key Laboratory of Computational Physics (Grant No. 6142A05230503).
The work of Z. Ma is supported by NSFC No. 12201401, No. 92270120 
and Beijing Institue of Applied Physics and Computational Mathematics funding HX02023-60.}}}

\author{Xizhe~Xie$^{1}$, Wengu~Chen$^{1,2}$, Zheng Ma$^{3}$, Han~Wang$^{1,2,4, *}$}

	
\maketitle

\begin{abstract}
The Gray Radiative Transfer Equations (GRTEs) are high-dimensional, multiscale problems that pose significant computational challenges for traditional numerical methods. 
Current deep learning approaches, including Physics-Informed Neural Networks (PINNs) and Asymptotically Preserving Neural Networks (APNNs), are largely restricted to low-dimensional or linear GRTEs. 
To address these challenges, we propose the Radiative Transfer Asymptotically Preserving Neural Network (RT-APNN), an innovative framework extending APNNs. RT-APNN integrates multiple neural networks into a cohesive architecture, reducing training time while ensuring high solution accuracy. 
Advanced techniques such as pre-training and Markov Chain Monte Carlo (MCMC) adaptive sampling are employed to tackle the complexities of long-term simulations and intricate boundary conditions. 
RT-APNN is the first deep learning method to successfully simulate the Marshak wave problem. 
Numerical experiments demonstrate its superiority over existing methods, including APNNs and MD-APNNs, in both accuracy and computational efficiency. 
Furthermore, RT-APNN excels at solving high-dimensional, nonlinear problems, underscoring its potential for diverse applications in science and engineering. 
\end{abstract}

\section{Introduction}
The Gray Radiative Transfer Equations (GRTEs) provide a mathematical framework for modeling thermal radiation transfer. These equations are widely used to analyze energy transmission and absorption within a medium, particularly in environments characterized by high temperatures and radiation intensities. GRTEs find applications in various critical fields, including astrophysics, nuclear engineering, and atmospheric science~\cite{chandrasekhar2013radiative,thomas2002radiative,howell2020thermal,clough2005atmospheric,pomraning2005equations}.
    
Achieving accurate simulations of the GRTEs presents substantial challenges, primarily due to two key factors: the high-dimensional and nonlinear nature of the radiation transport and material energy equations, and the multiscale characteristics of the opacity in background materials~\cite{densmore2004asymptotic,morel1996linear,smedley2015asymptotic}.

Numerical methods for solving GRTEs can be broadly categorized into two main approaches. The first category encompasses deterministic methods, such as the Pn method, which utilizes spherical harmonics expansion, and the discrete-ordinate Sn method~\cite{schafer2011diffusive,carlson1955solution,lathrop1964discrete,walters1991investigation,dahl1999positive}. 
In the case of GRTEs with a very small Knudsen number, the material exhibits high opacity, the photon mean free path decreases, and the radiative transfer equation approximates a diffusion equation as the Knudsen number approaches zero. In numerical simulations of GRTEs, the spatial mesh size is typically set to be on the order of the photon mean free path. As a result, simulations in the diffusion regime become computationally expensive. To mitigate this challenge, Sun et al.~\cite{sun2015asymptotic} introduced an Asymptotic-Preserving (AP) scheme. This scheme automatically satisfies the diffusion limit equation as the Knudsen number tends to zero, while maintaining effectiveness for larger Knudsen numbers.
Therefore, the Pn and Sn methods are typically integrated with asymptotically preserving numerical schemes, which have demonstrated potential in capturing multiscale features and have produced successful outcomes for specific problems~\cite{li2020unified, sun2015asymptotic, xu2020asymptotic, fu2022asymptotic, li2024asymptotic, xiong2022high, FRANK20072289, MCCLARREN20082864, MCCLARREN20087561, SUN2017455, XU20171}.
However, both the Pn and Sn methods can face difficulties in accurately solving problems involving complex geometries, highly absorptive media, or non-uniform angular distributions. 
Furthermore, in high-dimensional scenarios, these deterministic methods often incur significant computational costs after discretization.
The second category consists of stochastic methods, such as Monte Carlo (MC) and Implicit Monte Carlo (IMC) methods~\cite{fleck1961calculation,FLECK1971313}. 
These methods can effectively handle complex geometries and material properties without suffering from the exponential increase in computational complexity often encountered in high-dimensional spaces~\cite{steinberg2022multi,mcclarren2009modified,GENTILE2001543,DENSMORE20111116,shi2020continuous,shi2023efficient}. 
Notably, Monte Carlo methods avoid exponential cost scaling with dimensionality, as their computational complexity grows linearly with the number of dimensions. However, these methods still face significant challenges, including high computational costs driven primarily by statistical noise, which requires extensive sampling to achieve accurate results. 
This noise-induced cost, along with slow convergence rates, remains a persistent obstacle~\cite{densmore2004asymptotic}.

Summarizing the above discussion, existing numerical methods each face their own limitations, prompting researchers to explore deep learning approaches as a potential solution to these challenges.
Among these, Physics-Informed Neural Networks (PINNs) have gained significant attention by incorporating prior physical knowledge, such as the governing PDEs, directly into the neural network training process~\cite{raissi2019physics}. Furthermore, by representing solutions through deep neural networks, which excel at modeling high-dimensional functions, PINNs are particularly well-suited for solving high-dimensional PDEs~\cite{yang2020physics, wang2021eigenvector, jagtap2020extended}. As mesh-free methods, PINNs offer the advantage of solving PDEs without the need for complex grid discretization, in contrast to traditional numerical approaches~\cite{gao2021phygeonet}.

S. Mishra et al. are the first to use PINNs to solve radiative transfer equations (RTEs) and achieve excellent results in some steady-state and transient linear cases~\cite{MISHRA2021107705}.
However, when addressing multiscale GRTEs, PINNs also encounter challenges. In PINNs, as the Knudsen number approaches zero, specific terms in the loss function diminish, leading to a decrease in accuracy in the asymptotic limit and causing the network to converge to a trivial solution.
Recently, Asymptotic-Preserving Neural Networks (APNNs) have been introduced, combining the AP scheme with PINNs~\cite{jin2023asymptotic,lu2022solving}. This approach constructs a loss function with asymptotic-preserving properties by decomposing the radiation intensity. The two primary decomposition techniques are the micro-macro decomposition~\cite{xu2010unified,mieussens2013asymptotic,jin2023asymptotic} and the even-odd decomposition~\cite{jin1998diffusive,jin2000uniformly,miller1991analysis,seibold2014starmap,jin2023asymptoticpreserving}. Both methods embed physical laws and asymptotic properties into the training process, ensuring that as the scale parameter approaches zero, the loss function smoothly transitions from a transport state to a diffusion limit. Consequently, both decomposition approaches are effective for solving transport equations across diffusion scales, although it remains unclear which method provides a distinct advantage in practice.

A key limitation of APNNs is the necessity to train and optimize multiple networks concurrently, which substantially increases memory requirements and prolongs training durations. Additionally, the interactions among these networks add significant complexity to the optimization process. Notably, while APNNs have shown success in solving linear GRTEs, they remain inadequate for effectively addressing the challenges associated with nonlinear GRTEs~\cite{jin2023asymptotic,jin2023asymptoticpreserving}.
	
A recent study proposed Model-Data Asymptotic-Preserving Neural Networks (MD-APNNs) to overcome the challenges of solving GRTEs by integrating supervising data from traditional numerical methods into the training process~\cite{Li2022AMA}. Numerical results demonstrate that MD-APNNs outperform both APNNs and purely data-driven networks in simulating nonlinear and non-stationary GRTEs. However, despite these advancements, MD-APNNs face difficulties in accurately modeling temperature-dependent opacity and in minimizing their reliance on supervising data.

In this paper, we propose the RT-APNN method for solving GRTEs, which enhances the accuracy of standard APNNs for linear and steady-state problems and addresses challenges in nonlinear and time-dependent scenarios.
The primary innovation of RT-APNN lies in the newly proposed micro-macro network architecture that aligns with the micro-macro decomposition. Additionally, the introduction of pre-training and adaptive sampling methods~\cite{guo2023pre,yu2023mcmc} addresses challenges associated with long timescales and  time-evolving  sharp interfaces.
Notably, our approach relies exclusively on physical model constraints, thereby eliminating the need for supervising data.
Moreover, it achieves an accuracy that is comparable to, or even better than, that of MD-APNNs.
	
The structure of this paper is as follows. Section 2 provides background information, describing the GRTEs and the micro-macro decomposition method. In Section 3, we present our proposed RT-APNN method in detail, including the improved network architecture, the pre-training strategy for addressing long-term dependencies, and the MCMC method with collocation-adaptive sampling. Section 4 compares our method with other state-of-the-art approaches through several numerical experiments, covering linear, nonlinear, and time-dependent cases, demonstrating the superiority of our approach in these scenarios. Finally, Section 5 concludes the paper.

\section{Preliminaries}
    
\subsection{The gray radiative transfer equations}
	
Consider the scaled form of the GRTEs in a bounded domain \(\mathbb{T} \times D \times S\), where \(\mathbb{T}\) represents the time domain, \(D \subset \mathbb{R}^d\) is the spatial domain, and \(S\) denotes the angular domain, specifically the \(d\)-dimensional unit sphere \(\mathbb{S}^{d-1}\), which describes the set of possible directions for radiation propagation.
The governing equation reads
\begin{equation}\label{eq:GRTEs}
    \left\{
    \begin{aligned}
        & \frac{\varepsilon^2}{c} \frac{\partial I}{\partial t}+\varepsilon \Omega \cdot \nabla I =\sigma \left(\frac{1}{4 \pi}acT^4-I\right),    \\
        & \varepsilon^2 C_v\frac{\partial T}{\partial t} = \sigma \left(\int_{S} I \text{d} \Omega -acT^4\right), \\
        & \mathcal{B} I = 0, \; \\
        & I(t=0, x, \Omega) = I_0(x, \Omega), \\
        & T(t=0, x) = T_0(x),
    \end{aligned}
    \right.
\end{equation}
where $I(t,x,\Omega)$ is the radiation intensity, time variable $t \in \mathbb{T}$, space variable $x \in D \subset \mathbb{R}^d$, and angular direction $\Omega \in S =  \mathbb{S}^{d-1}$ i.e. the d-dimensional sphere, $T(t,x)$ is the material temperature, $a, c, \sigma$ denote the  radiation constant, scaled speed of light, opacity and $C_v$ is the scaled heat capacity. $ \mathcal{B}$ is the boundary operator for $I$. The parameter $\varepsilon > 0$ is called the Knudsen number which characterizes the ratio of mean free path over the characteristic length of the system. 
In the one-dimensional context, the GRTEs take the following form:
\begin{equation}\label{eq:GRTEs1d}
    \left\{
    \begin{aligned}
        & \frac{\varepsilon^2}{c} \frac{\partial I}{\partial t}+\varepsilon \mu \frac{\partial I}{\partial x} =\sigma \left(\frac{1}{2}acT^4 -I\right), \quad (t,x,\mu) \in \mathbb{T} \times D \times [-1,1],               \\
        & \varepsilon^2 C_v\frac{\partial T}{\partial t} = \sigma \left(\int_{-1}^{1}I \text{d} \mu -acT^4 \right), \quad (t,x) \in \mathbb{T} \times D.
    \end{aligned}
    \right.
\end{equation}

When the temperature of the material aligns with the temperature of radiation, expressed as $T_r = (\frac{1}{ac}\int I \mathrm{d}\Omega )^{1/4}$, eq.\eqref{eq:GRTEs} simplifies to the linear transport model
\begin{equation}
    \frac{\varepsilon^2}{c}\partial_t I + \varepsilon \Omega \cdot \nabla I = \sigma\left(\frac{1}{4\pi}\int_{S} I \text{d} \Omega -I\right).
\end{equation}
In 1D case, one can obtain
\begin{equation}
    \frac{\varepsilon}{c} \partial_t I+ \mu \partial_x I =\frac{\sigma}{\varepsilon} \left(\frac{1}{2}\int_{-1}^{1}I \text{d} \mu -I\right).
\end{equation}

Eq.\eqref{eq:GRTEs1d} represents a relaxation model pertaining to the radiation intensity within the context of local thermodynamic equilibrium, with the emission source originating from the background medium, as dictated by the Planck function corresponding to the local material temperature, more precisely, $\sigma acT^4 / {4\pi}$.
As the parameter $\epsilon$ tends towards zero, while disregarding boundaries and initial moments, the radiation intensity denoted as $I$ converges towards a Planck function at the local temperature. This can be stated as $I^{(0)}= ac\left(T^{(0)}\right)^4 / 4 \pi$. Additionally, the local temperature $T^{(0)}$ satisfies a diffusion equation:
\begin{equation}\label{eq:diffusion}
    \frac{\partial}{\partial t}\left(C_v T^{(0)}\right) + a \frac{\partial}{\partial t}\left(T^{(0)}\right)^4 = \nabla \cdot \frac{ac}{3\sigma}
    \nabla \left(T^{(0)}\right)^4.
\end{equation}

\subsection{Micro-macro decomposition}

The micro-macro decomposition is one of the decomposition methods used in APNNs~\cite{xu2010unified,mieussens2013asymptotic,jin2023asymptotic}.
We take the following form of micro-macro  decomposition used in~\cite{xiong2022high}. 
The radiation intensity \(I(t, x, \Omega)\) is decomposed into its equilibrium part \(\rho(t, x)\) and non-equilibrium part \(g(t, x, \Omega)\):
\begin{equation}\label{eq:micro-macro}
\begin{aligned}
    \left\{
        \begin{aligned}
    I(t, x, \Omega)&=\rho(t, x)+\frac{\varepsilon}{\sqrt{\sigma_0}} g(t, x, \Omega),\\
    \rho(t,x)&=\langle I(t,x,\Omega)\rangle,
        \end{aligned}
    \right.
\end{aligned}
\end{equation}
where \(\varepsilon\) is the micro-scale parameter, and \(\sigma_0>0\) is a constant defined as the reference opacity. 
The operator \(\langle\cdot \rangle\) is defined as $\langle f (\cdot,\Omega)\rangle=\frac{1}{|S|} \int f (\cdot,\Omega) \text{d} \Omega$, clearly \(\langle g(t,x,\Omega)\rangle=0\).
Through the decomposition, we can convert the original GRTE~\eqref{eq:GRTEs} into the following coupled system:
\begin{equation}\label{eq:loss}
\left\{
\begin{aligned}
    &\frac{1}{c} \partial_t \rho + \frac{1}{\sqrt{\sigma_0}}\nabla \cdot \left\langle \Omega g\right\rangle + \frac{1}{|S|} C_v \partial_t T = 0 ,\\
    &\frac{\varepsilon^2}{c} \partial_t g + \varepsilon\left(\nabla \cdot \left( \Omega g\right) - \nabla \cdot \left\langle \Omega g\right\rangle\right) + \sqrt{\sigma_0} \nabla \cdot \left( \Omega  \rho\right) + \sigma g = 0,\\
    &\varepsilon^2 C_v \partial_t T = \sigma\left(|S| \rho - a c T^4\right).
\end{aligned}
\right.
\end{equation}
When the micro-scale parameter \(\varepsilon \rightarrow 0\), the system's behavior converges to the asymptotic limit equation:
\begin{equation}\label{eq:diffusion_limit}
\left\{
\begin{aligned}
  &\frac{1}{c} \partial_t \rho + \frac{1}{\sqrt{\sigma_0}}\nabla \cdot \left\langle \Omega g\right\rangle + \frac{1}{|S|} C_v \partial_t T = 0 ,\\
  &\sqrt{\sigma_0} \nabla \cdot \left( \Omega  \rho\right) + \sigma g = 0,\\
  &0 = \sigma\left(|S| \rho - a c T^4\right).
\end{aligned}
\right.
\end{equation}
Solving the above equations jointly yields the nonlinear diffusion limit equation~\eqref{eq:diffusion}.

During the micro-macro decomposition process, it is essential to consider the boundary and initial conditions, which can be directly obtained from \eqref{eq:GRTEs} and \eqref{eq:micro-macro}. 
    \begin{equation}
    \left\{
    \begin{aligned}
        & \mathcal{B} (\rho +\frac{\epsilon}{\sqrt{\sigma_{0}}} g) = 0, \; \\
        & \rho(t=0, x, \Omega) +\frac{\epsilon}{\sqrt{\sigma_{0}}} g(t=0,x,\Omega)  = I_0(x, \Omega), \\
        & T(t=0, x) = T_0(x),
    \end{aligned}
    \right.
\end{equation}
where boundary conditions include inflow boundary conditions, Dirichlet boundary conditions, reflective boundary conditions, and periodic boundary conditions. The initial conditions represent the state of the system at the initial time. Specific details will be provided directly in the experimental section.

\subsection{Residual Neural Network framework}
The ResNet~\cite{he2016deep} is a commonly used deep neural network architecture that can be employed in PINNs to parameterize solutions~\cite{cheng2021deep}, particularly when the solutions exhibit complex nonlinear characteristics.
A $N$-layer ResNet $R^N_\theta(\mathbf{x}) $ is defined by the following equations:
\begin{equation}
\begin{aligned}\label{eq:resnet}
    h_{\theta}^{0}(\mathbf{x}) & = \mathbf{W}^{0} \mathbf{x} + \mathbf{b}^{0}, \\
    h_{\theta}^{n}(\mathbf{x}) & = h_{\theta}^{n-1}(\mathbf{x}) + \mathcal{F}_{\theta}^{n}(h_{\theta}^{n-1}(\mathbf{x})), \quad 1 \leq n \leq N-1, \\
    R^N_{\theta}(\mathbf{x}) & = \mathbf{W}^{N-1} h_{\theta}^{N-1}(\mathbf{x}) + \mathbf{b}^{N-1}.
\end{aligned}
\end{equation}
In Eq.~\eqref{eq:resnet}, $h_{\theta}^{n}(\mathbf{x})$ represents the $n^{th}$ hidden layer, $\mathcal{F}_{\theta}^{n}(\cdot)$ denotes the mapping function of the $n^{th}$ residual block, $\mathbf{W}^{n}$ and $\mathbf{b}^{n}$ correspond to the weight and bias parameters of the $n^{th}$ layer, respectively, and $\theta$ symbolizes all network parameters. 
Within ResNet, the mapping function of a residual block of size $l$ is typically defined as:
\begin{equation}
\mathcal{F}^n_{\theta}(\mathbf{x}) =\mathcal{F}^n(\mathbf{x}, \{\mathbf{W}^n_i,\mathbf{b^n}_i\}) = \sigma(\mathbf{W}^n_l \cdot \sigma(\mathbf{W}^n_{l-1} \cdot \sigma(\dots \sigma(\mathbf{W}^n_1 \cdot \mathbf{x} + \mathbf{b}^n_1) \dots) + \mathbf{b}^n_{l-1}) + \mathbf{b}^n_l),
\end{equation}
where $\sigma$ is a nonlinear activation function, $\mathbf{W}_i$  and $\mathbf{b}_i$ represents the weight matrix and the bias term of residual block.

\subsection{APNNs for GRTEs}

APNNs incorporates the asymptotic-preserving scheme into the PINNs, and defines three separate networks to model the micro-macro decomposed physical properties, $\rho, T, g$, as follows:
\begin{equation}
\begin{aligned}
\rho(t, x) & :=\rho_{\theta}(t, x)=\sigma^{+}\left[R_{\theta_\rho}(t, x)\right], \\
T(t, x) & :=T_{\theta}(t, x)=\sigma^{+}\left[R_{\theta_T}(t, x)\right], \\
g(t, x, \Omega) & :=g_{\theta}(t, x, \Omega)=R_{\theta_g}(t, x, \Omega),
\end{aligned}
\end{equation}
where $\sigma^+$ is a non-negative nonlinear function, ensuring that $\rho$ and $T$ do not take negative values. In practice, functions such as $e^{-x}$, $\ln(1+e^x)$, or the $\frac{1}{1+e^{-x}}$ may be selected.
The APNNs constructs an asymptotic-preserving loss function using micro-macro decomposition:
\begin{equation}
\mathcal{L}^{\varepsilon}(\theta)=w_r \mathcal{L}_r^{\varepsilon}(\theta)+w_i \mathcal{L}_i^{\varepsilon}(\theta)+w_b \mathcal{L}_b^{\varepsilon}(\theta),
\end{equation}
where the residual, initial condition and boundary condition losses are defined respectively by
\begin{equation}\label{eq:loss-terms}
\begin{aligned}
\mathcal{L}_r^{\varepsilon}(\theta) = & \frac{1}{N_r} \sum_{j=1}^{N_r} \Bigg\{ \bigg| \frac{1}{c} \partial_t \rho_{\theta }(t_j^r, x_j^r) + \frac{1}{\sqrt{\sigma_0}} \nabla \cdot \langle \Omega  g_{\theta}(t_j^r, x_j^r, \Omega_j^r) \rangle + \frac{1}{|S|} C_v \partial_t T_{\theta}(t_j^r, x_j^r) \bigg|^2 \\
& + \bigg| \varepsilon^2 C_v \partial_t T_{\theta}(t_j^r, x_j^r) - \sigma(|S| \rho_{\theta}(t_j^r, x_j^r) - a c T_{\theta}(t_j^r, x_j^r)^4) \bigg|^2 \\
& + \bigg| \varepsilon \nabla \cdot\big( \Omega  g_{\theta}(t_j^r, x_j^r, \Omega_j^r) - \nabla \cdot\langle \Omega  g_{\theta}(t_j^r, x_j^r, \Omega_j^r) \rangle \big) + \frac{\varepsilon^2}{c} \partial_t g_{\theta}(t_j^r, x_j^r, \Omega_j^r) \\
& + \sqrt{\sigma_0} \nabla \cdot(\Omega \rho_{\theta}(t_j^r, x_j^r)) + \sigma g_{\theta}(t_j^r, x_j^r, \Omega_j^r) \bigg|^2 \Bigg\},\\
\mathcal{L}_{i}^{\varepsilon}(\theta)  =&\frac{1}{N_i}\sum_{j=1}^{N_i}\left\{\bigg|\rho_\theta\left(0, x_j^i\right)+\frac{\varepsilon}{\sqrt{\sigma_0}} g_\theta\left(0, x_j^i, \Omega_j^i\right)-I_0\left(x_j^i, \Omega_j^i\right)\bigg|^2+\bigg|T_\theta\left(0, x_j^i\right)-T_0\left(x_j^i\right)\bigg|^2 \right\},\\
\mathcal{L}_{b}^{\varepsilon}(\theta)  =&\frac{1}{N_{b}} \sum_{j=1}^{N_{b}}\left|\mathcal{B}\left(\rho_\theta(t_j^b,x_j^b)+\frac{\varepsilon}{\sqrt{\sigma_0}} g_\theta(t_j^b,x_j^b,\Omega_j^b)\right)\right|^2.
\end{aligned}
\end{equation}

To verify the asymptotic-preserving property, we set the parameter $\varepsilon \rightarrow 0$, thereby eliminating terms containing $\varepsilon$. This yields the following loss function for the asymptotic-preserving formulation:
\begin{equation}
\begin{aligned}
\mathcal{L}_r(\theta) = & \frac{1}{N_r} \sum_{j=1}^{N_r} \left\{ \left| \frac{1}{c} \partial_t \rho_\theta(t_j^r, x_j^r) + \frac{1}{\sqrt{\sigma_0}} \nabla \cdot\langle \Omega  g_\theta(t_j^r, x_j^r, \Omega_j^r) \rangle + \frac{1}{|S|} C_v \partial_t T_\theta(t_j^r, x_j^r) \right|^2 \right. \\
& + \left| \sigma \left(|S| \rho_\theta(t_j^r, x_j^r) - a c T_\theta(t_j^r, x_j^r)^4 \right) \right|^2 + \left| \sqrt{\sigma_0} \nabla \cdot(\Omega  \rho_\theta(t_j^r, x_j^r)) + \sigma g_\theta(t_j^r, x_j^r, \Omega_j^r) \right|^2 \Bigg\}.
\end{aligned}
\end{equation}
This loss function precisely corresponds to the asymptotic limit equation~\eqref{eq:diffusion_limit}.

\section{Radiative Transfer Asymptotically Preserving Neural Networks (RT-APNN)}

The RT-APNN method is developed from the APNNs method by incorporating the micro-macro network architecture, pre-training strategy, and adaptive sampling techniques. The motivation behind the micro-macro network architecture stems from the fact that the variables in the system of equations are coupled, and we aim to avoid using two separate neural networks to model them. A natural approach is to couple the two networks together using concatenation techniques. 

The pre-training strategy is inspired by traditional time-stepping methods commonly used in numerical approaches for solving long-time problems, which enhances the stability and accuracy of the solution process. The purpose of employing the MCMC adaptive sampling technique is to intelligently adjust the distribution of collocation points, enabling the use of fewer points to solve the problem and thus reducing memory requirements.

These ideas collectively allow the RT-APNN method to improve both the efficiency and accuracy of solving radiative transfer equations, particularly in complex scenarios involving nonlinearity and high-dimensional spaces. The following sub-sections will provide a detailed explanation of each of these methods.

\subsection{The micro-macro network structure of RT-APNN}

\begin{figure}[]
    \centering
    \includegraphics[width=0.9\textwidth]{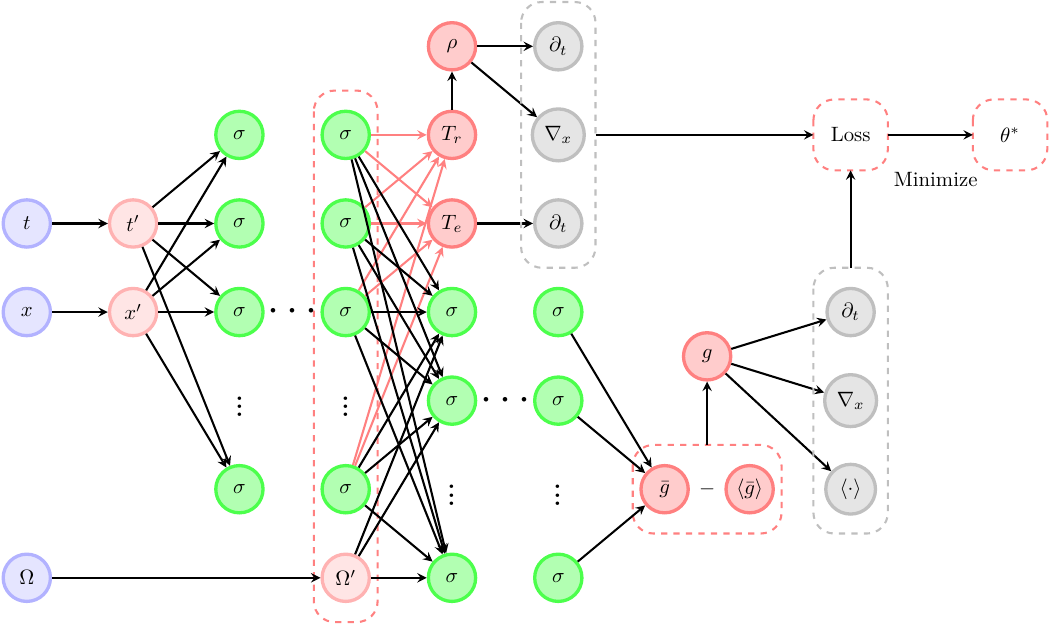}
    \caption{\@ Micro-macro network structure of RT-APNN for solving the GRTEs model.}\label{fig:net}
\end{figure}

The micro-macro network structure of RT-APNN is depicted in Figure~\ref{fig:net}.
Initially, we preprocess the input variables \( t \), \( x \), and \( \Omega \) using a scaling function \( L(x) \) to map each component element-wise into the range \([-1, 1]\) by
\begin{equation}\label{eq:scale}
y^{\prime} = L(y) = 2 \frac{(y - y_{\min})}{y_{\max} - y_{\min}} - 1, \quad y \in [y_{\min}, y_{\max}],
\end{equation}
In Eq.~\eqref{eq:scale}, $y$ takes input variables \( t \), \( x \), and \( \Omega \) and defines scaled inputs  \( t' \), \( x' \), and \( \Omega' \), respectively.
This preprocessing step facilitates faster convergence of the gradient descent algorithm.
Subsequently, the temporal and spacial variables \( (t^{\prime}, x^{\prime}) \) are processed through an \( N \)-layer residual network to obtain two macro-scale outputs: radiation temperature \( T_r \) and material temperature \( T_e \):
\begin{equation}\label{eq:trte}
(T_r, T_e)^T = \sigma^{+}[R_{\theta_T}(t^{\prime}, x^{\prime})] .
\end{equation}
Each residual block within the network comprises two sub-layers. Notably, the radiation temperature \( T_r \) directly yields 
\begin{equation}
    \rho = \frac{1}{|S|} a c T_r^4.
\end{equation}
According to Eq.~\eqref{eq:trte}, the last hidden layer \( h_{\theta_T}^{N-1} \) of \(R_{\theta_T}\) encodes features representing the macroscopic temperature variables, which are further processed to represent the microscopic variable \( g \). 
We concatenate the angular variable \( \Omega^{\prime} \) with the output of the hidden layer \( h_{\theta_T}^{N-1}(t^{\prime}, x^{\prime}) \) to form a new hidden layer.
This new hidden layer is then passed through an additional residual network, resulting in the output \( \bar{g} \):
\begin{equation}
\bar{g} = R_{\theta_g}(h_{\theta_T}^{N-1}(t^{\prime}, x^{\prime}), \Omega^{\prime}).
\end{equation}
Each residual block of \( R_{\theta_g} \) in this network comprises two sub-layers. Additionally, \( R_{\theta_g} \) may have a different number of layers compared to \( N \).
Finally, in the model, we define
\begin{equation}
   g = \bar{g} - \langle \bar{g} \rangle
\end{equation}
that ensures the equilibrium condition \( \langle g \rangle = 0 \) is consistently satisfied. The average \( \langle \cdot \rangle \) is approximated via numerical integration:
\begin{equation}
\langle \bar{g} \rangle \approx \frac{1}{|S|} \sum_i w_i R_{\theta_g}(h_{\theta_T}^{N-1}(t^{\prime}, x^{\prime}), L(\Omega_i)).
\end{equation}
In the one-dimensional case, \( \Omega_i \) represents Gaussian integration points and \( w_i \) denotes Gaussian integration weights.
In the two-dimensional case, \( \Omega_i \) signifies Lebedev points and \( w_i \) denotes Lebedev weights.

\subsection{Pre-training strategy for solving evolution equations}

The pre-training method proposed for time evolution equations~\cite{guo2023pre} decomposes the time interval $\mathbb{T}$ into multiple sub-intervals and sequentially trains them in chronological order:
\begin{equation}
[0, T_1], [0, T_2], \cdots, [0, T_{n}], \quad T_1<T_2<\cdots<T_{n}=T.
\end{equation}
Initially, a complete training is conducted within the interval $[0, T_1]$, yielding trained network parameters $\theta_1^*$. 
Sampling $N_1$ points from the domain contained within $[0, T_1]$, denoted as $V_1 = \{(t_j, x_j) | t_j \in [0, T_1], 1 \leq j \leq N_1\}$, and then utilizing the trained network to predict the solutions $U_1$ on $V_1$. 
Specifically, in our experiments, the predicted solution \( U_1 \) is the union of the radiation temperature \( \{ Tr_{\theta_1^*}(t_j, x_j) \mid (t_j, x_j) \in V_1 \} \) and the material temperature \( \{ Te_{\theta_1^*}(t_j, x_j) \mid (t_j, x_j) \in V_1 \} \).

\begin{figure}[]
    \centering
    \includegraphics[width=0.9\textwidth]{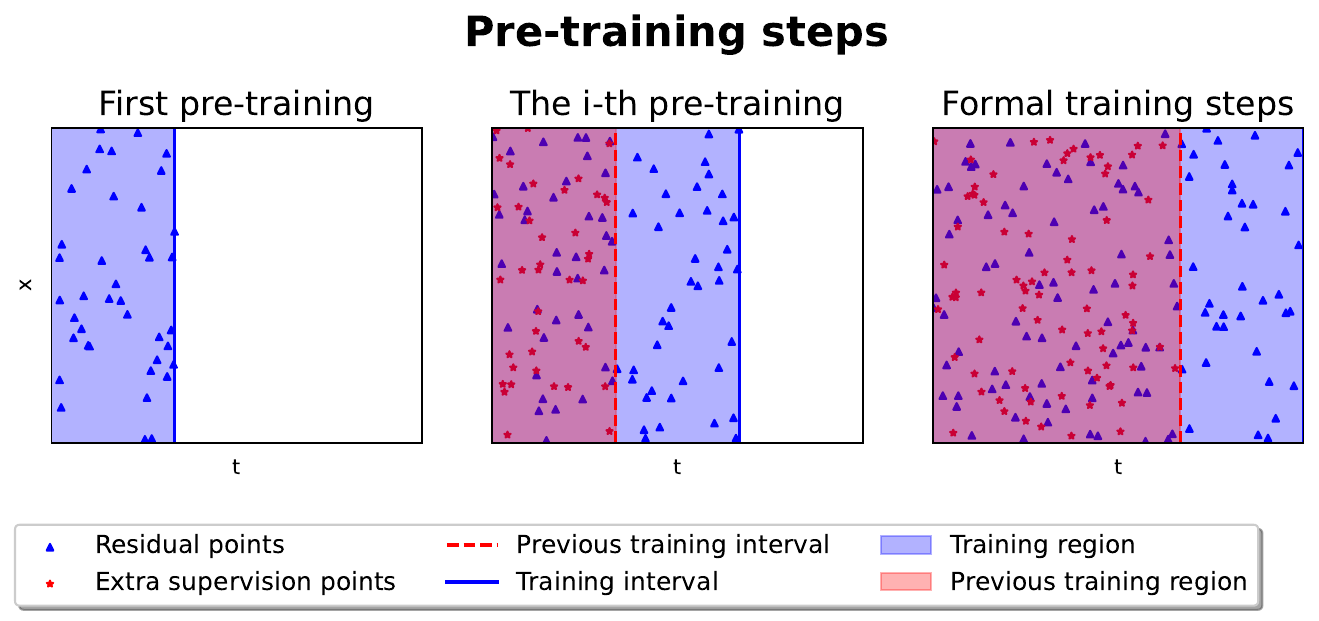}
    \caption{\@ Example of a pretraining strategy for solving time-dependent PDEs.}\label{fig:pre-training}
\end{figure}

Subsequently, when training within the interval $[0, T_2]$, the saved network parameters $\theta_1^*$ are used as initial parameters of the neural networks, and the following optimization problem is  solved:
\begin{equation}
\theta_2^{*}=\arg \min_{\theta_2}\bigg\{ \mathcal{L}\left(\theta_2 ;  \theta_1^{*}\right)+\omega_s\mathcal{L}_s\left(\theta_2 ;  \theta_1^{*},U_1\right)\bigg\},
\end{equation}
where $\mathcal{L}(\theta_2; \theta_1^{*})$ represents the loss formed by the network parameters $\theta_2$ initialized with $\theta_1^{*}$. 
The additional pseudo-supervising loss term $\mathcal L_s$ is introduced to ensure that the previously trained regions do not undergo significant changes when training progresses to subsequent intervals.
\begin{equation}
\mathcal{L}_s\left(\theta_2 ;  \theta_1^{*},U_1\right)=\frac{1}{N_{1}} \sum_{j=1}^{N_{1}}\bigg(\bigg|Tr_{\theta_2}\left(t_j,x_j\right)-Tr_{\theta_1^*}\left(t_j,x_j\right)\bigg|^2 + \bigg|Te_{\theta_2}\left(t_j,x_j\right)-Te_{\theta_1^*}\left(t_j,x_j\right)\bigg|^2 \bigg).
\end{equation}
As $U_1$ is not a reference solution, it is referred to as a pseudo-label. 
$\omega_s$ denotes the weight of the pseudo-supervising loss, and when $\omega_s$ set to 0, this term is  ignored.
The training process for subsequent intervals is repeated in a similar manner until the entire time domain is fully trained. Examples of the training process can be found in Figure~\ref{fig:pre-training}.

\subsection{MCMC adaptive sampling strategy}

The core concept of the Markov Chain Monte Carlo (MCMC) adaptive sampling method~\cite{yu2023mcmc} is to optimally distribute the collocation points used in the loss function~\eqref{eq:loss-terms} based on the error distribution of the solution. Regions with larger errors typically require more sampling points, and the goal of the optimal distribution is to minimize the solution error within a fixed budget \(N_r\) of sampling points.

The first step in MCMC adaptive sampling is to define the desired probability distribution for the sampling points. The error distribution of the solution is commonly used as an indicator function for this distribution. However, directly quantifying the error can be challenging.

In this work, we use the residuals as the indicator function, where regions with larger residuals are assigned higher sampling probabilities. The probability distribution is defined as follows: 
\begin{equation}\label{eq:dist}
\pi(t, x):=\log _{10}\left(1+
\frac 1\alpha {\Big\vert \varepsilon^2 C_v \partial_t T-\sigma\left(|S| \rho-a c T^4\right) \Big\vert}\right),
\end{equation}
where \(\alpha\) is a constant used to prevent the indicator function from becoming too small, with \(\alpha = 10^{-16}\) chosen for our experiments.

Subsequently, we employ the MCMC method to sample collocation points according to the distribution defined in \eqref{eq:dist}. In our experiment, the proposal for the update is defined as follows:
\begin{equation}
\begin{aligned}
x' &= U(x) = x + \epsilon \Delta x, \quad \Delta x \sim N(0,1^2), \\
x^* &= U_b - \left| U_b - L_b -\mathrm{mod}(x' - L_b, 2U_b - 2L_b) \right|,
\end{aligned}
\end{equation}
where \(\epsilon\) is the step size of the random walk, and \(U_b\) and \(L_b\) represent the upper and lower bounds of the spatio-temporal domain, respectively. The update function is designed to ensure that points moving outside the computational domain are redirected back into the interior. After several iterations, the final sample points will approximate the desired distribution.

It is important to note that the residual-based distribution in \eqref{eq:dist} serves as an unnormalized probability density function, as its normalization constant is unknown, making direct sampling from this distribution infeasible. The MCMC adaptive sampling method significantly reduces the number of required points while enhancing the accuracy of the numerical solution. The detailed MCMC procedure is outlined in Algorithm~\ref{alg:MCMC}.

While the algorithm conceptually involves two loops, practical implementation can achieve the same results using only a single loop. This optimization is possible because the sampling points can be represented as tensors, which enables parallel operations. Such an approach can significantly improve computational efficiency.
\begin{algorithm}
    \caption{Adaptive sampling in MCMC~\cite{yu2023mcmc}.}
    \label{alg:MCMC}
    \begin{algorithmic}[1]
        \REQUIRE Initial points $\{( t_j^{(0)},x_j^{(0)})\}_{j=1}^{N_{\text{r}}}$, upper bound $U_b$, lower bound $L_b$, Update function $U$,Indicator function $\pi$,
        step size $\epsilon$, steps $N_{\text{mcmc}}$.
        \FOR{$j=1$ to $N_r$}
        \FOR{$i=1$ to $N_{\text{mcmc}}$}
        \STATE Generate $u$ from $\mathcal{U}[0,1]$.
        \STATE Sample new intermediate state $( t',x') \sim \mathcal{N}(( t_j^{(i-1)},x_j^{(i-1)}), \Sigma)$, 
        \STATE Map to new proposal state: $( t^*,x^*)=U( t',x')$.
        \STATE Calculate acceptance probability:
        \begin{equation}
        \alpha\left(t_j^{(i-1)}, x_j^{(i-1)} ; t^*, x^*\right)=\min \left\{1, \frac{\pi\left(t_j^{(i-1)}, x_j^{(i-1)}\right) }{\pi\left(t^*, x^*\right) }\right\}
        \end{equation}
        \IF{$u < \alpha(t_j^{(i-1)}, x_j^{(i-1)}; t^*, x^*)$}
        \STATE Accept proposal: $(t_j^{(i)}, x_j^{(i)})=(t^*, x^*)$.
        \ELSE
        \STATE Reject proposal: $(t_j^{(i)}, x_j^{(i)})=(t_j^{(i-1)}, x_j^{(i-1)})$.
        \ENDIF
        
        \ENDFOR
        \STATE Set $(t_j, x_j)=(t_j^{(N_{\text{mcmc}})}, x_j^{(N_{\text{mcmc}})})$.
        \ENDFOR
        \ENSURE The newly obtained collocation points 
    \end{algorithmic}
\end{algorithm}

\section{Experiments}

In this section, we conduct 6 numerical experiments to validate the performance of the proposed RT-APNN method for GRTEs. 
We provide several examples to compare the performance of PINNs, APNNs, MD-APNNs, and RT-APNN. Due to the integral terms in the loss function for the system \eqref{eq:loss}, we approximate them using a 10-point Gauss-Legendre quadrature rule for one-dimensional problems and 50 Lebedev quadrature points for two-dimensional problems.
The effect of the number of quadrature nodes and sampling points on solution accuracy is presented in Appendix~\ref{app:a}.
The reference solutions are obtained through the spherical harmonics method $P_{12}$. We use the relative $L^2$ error as the evaluation criterion, defined as follows:
\begin{equation}
L_{\text{error}}^2(u) = \frac{\left\|u_{nn} - u_{\textrm{ref}}\right\|_{L^2}}{\left\|u_{\textrm{ref}}\right\|_{L^2}},
\end{equation}
where \( u \) represents the radiation temperature \( T_r \) or the material temperature \( T_e \).

\subsection{Experimental setup}
In all the experiments conducted, we use a ResNet architecture with a width of 64. 
For the linear transport equation, the network's output includes the macroscopic component $\rho$ and the microscopic component $g$.
For the coupled nonlinear radiation transport equations with material temperature, the network's output comprises the radiation temperature $T_r$, material temperature $T_e$, and $g$.
In the micro-macro-net, the part that outputs $\rho$ or $T_r$ and $T_e$ consists of two residual blocks, while the part that outputs $g$ consists of one residual block.
The activation function $\sigma(x)$ chosen is the Gaussian Error Linear Unit (GELU)~\cite{hendrycks2016gaussian}. 

The process of training the network is divided into two stages.
In the first stage, we use the Adam optimizer based on gradient descent~\cite{kingma2014adam}.
Unless specified otherwise, the initial learning rate is 0.01. 
We use a piecewise constant learning rate decay strategy, where the learning rate decreases by $5\%$ every 100 iterations, with a minimum learning rate of \(10^{-6}\). 
The second stage uses the optimization algorithm L-BFGS until the stopping criterion is met~\cite{liu1989limited}. 
We use Latin hypercube sampling for the spatiotemporal domain and uniform sampling on the sphere for the velocity direction, then concatenate these two parts. 
During training, MCMC adaptive sampling techniques are employed to adjust the spatial distribution of collocation points.
The number of iterations for MCMC is 10, with a step size of 1. 
The pre-training time intervals, the number of Adam iterations, and the number of collocation points for each experiment are summarized in Table~\ref{tab:Hyperparameter Settings}.
In the ablation studies, we use \ding{172} to indicate the use of a micro-macro-network, \ding{173} to indicate the use of adaptive sampling, and \ding{174} to indicate the use of a pre-training strategy.
For example, APNNs+\ding{172}\ding{173}\ding{174} denotes  that we use the micro-macro-network, adaptive sampling and pre-training strategies simultaneously, which gives the  RT-APNN method.

\begin{table}[]
    \caption{Summary Table of Hyperparameter Settings}
    \label{tab:Hyperparameter Settings}
    \centering
    \begin{tabular}{cccccc}
        \toprule
        \textbf{Example} & \textbf{Pre-training Times} & \textbf{Adam Iterations} & \textbf{$N_r$} & \textbf{$N_b$}& \textbf{$N_i$}\\
        \midrule
        Ex 1 & [1.0, 2.0] & 10000 & 16384 & 4096 & 4096 \\
        Ex 2 & [0.3, 0.6, 1.0] & 10000 & 16384 & 4096 & 4096 \\
        Ex 3 & [0.1, 0.3, 0.5] & 10000 & 16384 & 4096 & 4096 \\
        Ex 4 & [0.5, 1.0] & 10000 & 16384 & 4096 & 4096 \\
        Ex 5 $(\sigma_0 = 30)$ & 0.1 + [0.1, 0.4, 0.7, 1.0] & 20000 & 16384 & 1024 & 3240 \\
        Ex 5 $(\sigma_0 = 300)$ & 1 + [15, 30, 45, 60, 75] & 20000 & 16384 & 1024 & 1600\\
        Ex 6 & [0.5, 1.0] & 50000 & 4096 & 4096 & 4096 \\
        \bottomrule
    \end{tabular}
\end{table}

\subsection{Solve the 1D radiation transport equation}

\subsubsection{Ex 1: Linear GRTEs with Constant Scattering Coefficients at Diffusion Scale}\label{sec:ex1}
We conduct the first experiment using a 1D linear radiation transport equation in a diffusion regime with $\varepsilon=10^{-8}$:
\begin{equation}
\left\{
\begin{aligned}
&\frac{\varepsilon^2}{c} \partial_t I+\varepsilon\mu \partial_x I=\sigma(\langle I\rangle-I), \quad (t,x, \mu) \in\mathbb{T}\times D \times[-1,1], \\ 
&I\left(t, x_L, \mu>0\right)=1,\\
&I\left(t, x_R, \mu<0\right)=0, \\
&I(0, x, \mu)=0.
\end{aligned}
\right.
\end{equation}
In this study, we set the parameters as follows: $c=1$, $\sigma=1$, $\mathbb{T}=[0,2]$, and $D=[0,1]$. Although the equation is a highly simplified linear transport problem, the extremely small value of $\varepsilon$ renders vanilla PINNs incapable of solving it effectively. 

\begin{figure}[]
    \centering
    \includegraphics[width=1.\textwidth]{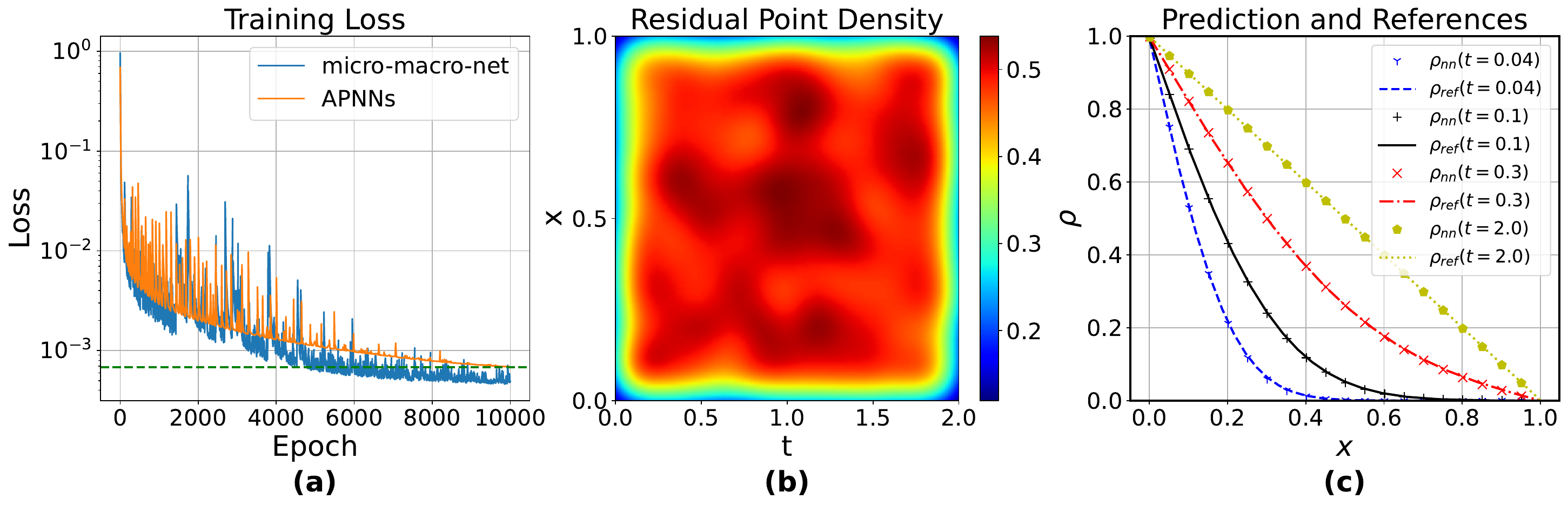}
    \caption{\@ The result of Ex 1. (a): The loss decay curves obtained by training with APNNs and APNNs+\ding{172}. (b): The distribution density function of the collocation points obtained by RT-APNN through MCMC adaptive sampling. (c): The radiation energy density $\rho$ at $t=0.04, 0.1, 0.3, 2.0$ obtained by RT-APNN for $\sigma = 1$.}\label{fig:ex1}
\end{figure}

We applied APNNs and RT-APNN to address this problem, with the corresponding loss curves shown in Figure~\ref{fig:ex1}(a). Notably, the micro-macro-net achieves comparable performance to APNNs in approximately 5,000 epochs, compared to 10,000 epochs required for APNNs. Figure~\ref{fig:ex1}(b) illustrates how MCMC sampling adaptively increases the density of collocation points in regions with higher residuals. Furthermore, Figure~\ref{fig:ex1}(c) presents the radiation energy density $\rho$ obtained using the RT-APNN method.

The relative errors are evaluated at time instances $t=0.04, 0.1, 0.3, 2.0$, as detailed in Table~\ref{tab:error-ex1}. The results indicate that the RT-APNN method improves accuracy by approximately half an order of magnitude compared to the APNN method.

\begin{small}
    \begin{table}[]
        \caption{The relative $L^2$ error of $\rho$ at time $t =0.04, 0.1, 0.3, 2.0$ for the diffusion regime ($\varepsilon = 10^{-8}$) in Ex 1 \cite{Li2022AMA}.}
        \label{tab:error-ex1}
        \centering
        \begin{tabular}{ccccc}
            \toprule[1pt]
            $L^2$ error & $t=0.04$ & $t=0.1$ & $t=0.3$ & $t=2.0$ \\
            \midrule[1pt] 
            PINNs & $7.05 \mathrm{e}-01$ & $7.57 \mathrm{e}-01$ & $7.49 \mathrm{e}-01$ & $3.63 \mathrm{e}-01$ \\
            APNNs & $4.54 \mathrm{e}-02$ & $1.62 \mathrm{e}-02$ & $5.31 \mathrm{e}-03$ & $5.78 \mathrm{e}-03$ \\
            
            RT-APNN & $1.61 \mathrm{e}-02$ & $4.73 \mathrm{e}-03$ & $3.64 \mathrm{e}-03$ & $1.55 \mathrm{e}-03$ \\
            \bottomrule[1pt]
        \end{tabular}
    \end{table}
\end{small}

We conduct a set of ablation experiments to compare the impact of different methods on the results. Since it is uncertain when the L-BFGS optimizer meets the stopping criteria, all methods were trained for 10,000 steps using the Adam optimizer.  
The results of these methods are presented in Table~\ref{tab:error-ex1-ablation}. Since the pretraining method involves stage-wise training, we allocated 10,000 steps for each training interval.  
Given the relatively simple nature of the problem, APNNs already demonstrate strong performance, leading to minimal differences in performance among the other methods.

\begin{small}
    \begin{table}[]
        \caption{The relative \(L^2\) error of \(\rho\) at times \(t = 0.04, 0.1, 0.3, 2.0\) with Adam training 10,000 steps in the diffusion regime (\(\varepsilon = 10^{-8}\)) in Ex 1.}
        \label{tab:error-ex1-ablation}
        \centering
        \begin{tabular}{lcccc}
            \toprule[1pt]
            $L^2$ error & $t=0.04$ & $t=0.1$ & $t=0.3$ & $t=2.0$ \\
            \midrule[1pt] 
            APNNs & $2.99 \mathrm{e}-02$ & $9.14 \mathrm{e}-03$ & $3.84 \mathrm{e}-03$ & $4.54 \mathrm{e}-04$ \\
            APNNs+\ding{172} & $1.24 \mathrm{e}-02$ & $6.17 \mathrm{e}-03$ & $4.02 \mathrm{e}-03$ & $2.98 \mathrm{e}-03$ \\
            APNNs+\ding{172}\ding{173} & $1.34 \mathrm{e}-02$ & $6.04 \mathrm{e}-03$ & $6.10 \mathrm{e}-03$ & $1.94 \mathrm{e}-03$ \\
            APNNs+\ding{172}\ding{173}\ding{174} (RT-APNN) & $1.55 \mathrm{e}-02$ & $6.89 \mathrm{e}-03$ & $2.33 \mathrm{e}-03$ & $9.89 \mathrm{e}-04$ \\
            \bottomrule[1pt]
        \end{tabular}
    \end{table}
\end{small}                                                                          

\subsubsection{Ex 2: Linear GRTEs with Variable Scattering Coefficients at Intermediate Scale}
We consider the radiative transport equation with a variable scattering coefficient \(\sigma = 1 + 10x^2\). The parameters are set as follows: \(\varepsilon = 10^{-2}\), \(c = 1\), \(\mathbb{T} = [0, 1]\), and \(D = [0, 1]\). 

\begin{figure}[]
    \centering
    \includegraphics[width=1.\textwidth]{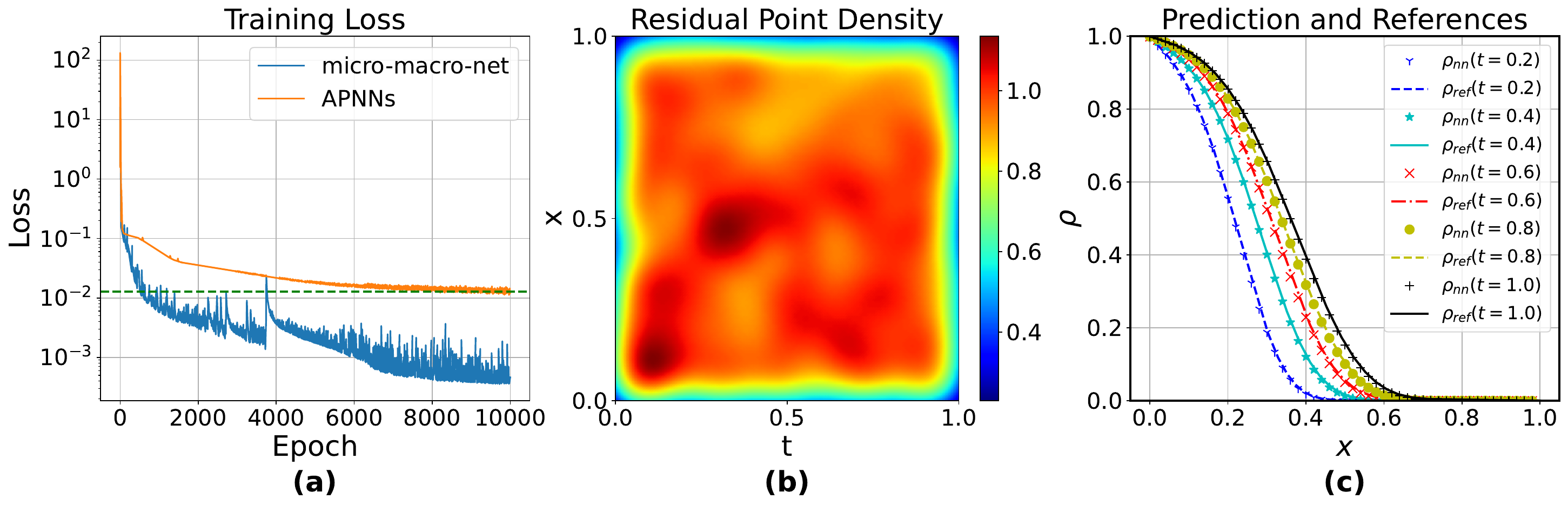}
    \caption{\@ The result of Ex 2. (a): The loss decay curves obtained by training with APNNs and APNNs+\ding{172}. (b): The distribution density function of the collocation points obtained by RT-APNN through MCMC adaptive sampling. (c): The radiation energy density $\rho$ at $t=0.2, 0.4, 0.6, 0.8, 1.0$ obtained by solving with RT-APNN for $\sigma = 1+10x^2$.}\label{fig:ex2}
\end{figure}

Figure~\ref{fig:ex2} presents the loss descent curves for APNNs and the micro-macro-net, along with the distribution of collocation points obtained using the MCMC adaptive sampling technique in RT-APNN. As shown in Figure~4(a), APNNs require 10,000 training epochs to reach a specific loss level, whereas RT-APNN achieves the same loss in fewer than 1,000 epochs. Figure~4(b) demonstrates that most collocation points are concentrated in the region \(x < 0.6\), and as time \(t\) increases, the points shift further back in space. This aligns with the solution in Figure~4(c), where \(\rho\) approaches zero for \(x > 0.6\). Furthermore, as \(t\) increases, the position where \(\rho\) vanishes also moves backward, indicating the success of the MCMC adaptive sampling technique in capturing this behavior.

Figure~\ref{fig:ex2}(c) presents the estimation of the radiative energy density \(\rho\) obtained using RT-APNN at the time instances \(t = 0.2, 0.4, 0.6, 0.8, 1.0\). The results are in close agreement with the reference solutions. Additionally, Table~\ref{tab:error-ex2} confirms that RT-APNN achieves the highest relative \(L^2\) accuracy among the methods compared.

We also conduct a series of ablation experiments using the Adam optimizer with 10,000 training steps. 
The results of the different methods are presented in Table~\ref{tab:error-ex2-ablation}. In contrast to the previous experiment, the micro-macro-net method demonstrates a significant improvement in accuracy, while the MCMC and pre-training strategies yield minimal enhancements.

\begin{small}
    \begin{table}[]
        \caption{The relative $L^2$ error of $\rho$ at time $t = 0.2,0.4,0.6,0.8,1.0$ for the intermediate regime ($\varepsilon = 10^{-2}$) in Ex 2  \cite{Li2022AMA}.}
        \label{tab:error-ex2}
        \centering
        \begin{tabular}{cccccc}
            \toprule[1pt]
            $L^2$ error & $t=0.2$ & $t=0.4$ & $t=0.6$ & $t=0.8$ & $t=1.0$ \\
            \midrule[1pt]  PINNs & $9.25 \mathrm{e}-01$ & $8.99 \mathrm{e}-01$ & $8.75 \mathrm{e}-01$ & $8.51 \mathrm{e}-01$ & $8.27 \mathrm{e}-01$ \\
            APNNs & $3.44 \mathrm{e}-02$ & $2.76 \mathrm{e}-02$ & $2.59 \mathrm{e}-02$ & $2.56 \mathrm{e}-02$ & $2.33 \mathrm{e}-02$ \\
            RT-APNN & $9.32 \mathrm{e}-03$ & $6.87 \mathrm{e}-03$ & $7.65 \mathrm{e}-03$ & $6.74 \mathrm{e}-03$ & $6.33 \mathrm{e}-03$ \\
            \bottomrule[1pt]
        \end{tabular}
    \end{table}
\end{small}

\begin{small}
    \begin{table}[]
        \caption{The relative $L^2$ error of $\rho$ at time $t = 0.2,0.4,0.6,0.8,1.0$ with Adam training 10,000 steps for the intermediate regime ($\varepsilon = 10^{-2}$) in Ex 2.}
        \label{tab:error-ex2-ablation}
        \centering
        \begin{tabular}{lccccc}
            \toprule[1pt]
            $L^2$ error & $t=0.2$ & $t=0.4$ & $t=0.6$ & $t=0.8$ & $t=1.0$ \\
            \midrule[1pt]  
            APNNs & $7.84 \mathrm{e}-02$ & $2.68 \mathrm{e}-02$ & $1.46 \mathrm{e}-02$ & $1.05 \mathrm{e}-02$ & $1.72 \mathrm{e}-02$ \\
            APNNs+\ding{172} & $1.29 \mathrm{e}-02$ & $6.93 \mathrm{e}-03$ & $7.02 \mathrm{e}-03$ & $6.14 \mathrm{e}-03$ & $5.98 \mathrm{e}-03$ \\
            APNNs+\ding{172}\ding{173} & $1.18 \mathrm{e}-02$ & $6.63 \mathrm{e}-03$ & $7.69 \mathrm{e}-03$ & $6.97 \mathrm{e}-03$ & $7.15 \mathrm{e}-03$ \\
            APNNs+\ding{172}\ding{173}\ding{174} (RT-APNN) & $9.26 \mathrm{e}-03$ & $6.64 \mathrm{e}-03$ & $6.62 \mathrm{e}-03$ & $5.64 \mathrm{e}-03$ & $5.57 \mathrm{e}-03$ \\
            \bottomrule[1pt]
        \end{tabular}
    \end{table}
\end{small}

\subsubsection{Ex 3: Nonlinear GRTEs with Periodic Boundary Conditions}\label{sec:ex3}

We address a 1D GRTEs problem with smooth initial conditions at equilibrium and periodic boundary conditions. 
\begin{equation}
    \left\{
    \begin{aligned}
        &\frac{\varepsilon^2}{c} \frac{\partial I}{\partial t}+\varepsilon \mu \frac{\partial I}{\partial x}=\sigma\left(\frac{1}{2} a c T^4-I\right),\quad(t, x, \mu) \in \mathbb{T} \times D \times[-1,1], \\
        &\varepsilon^2 C_v \frac{\partial T}{\partial t}=\sigma\left(2\langle I\rangle-a c T^4\right),\quad (t, x) \in \mathbb{T} \times D, \\
        &I\left(t, x_L, \mu\right)=I\left(t, x_R, \mu\right), \\
        &I(0, x, \mu)=\frac{1}{2} a c T(0, x)^4, \quad T(0, x)=\frac{3+\sin (\pi x)}{4}.
    \end{aligned}
    \right.
\end{equation}
The spatial domain \(D\) spans \([0, 2]\), the time interval \(\mathbb{T}\) is \([0, 0.5]\), and the parameters are set as \(a = c = 1\), \(C_v = 0.1\), and \(\sigma = 10\). We consider two cases with the kinetic scale \(\varepsilon = 1\) and \(\varepsilon = 10^{-3}\).

\begin{figure}[]
    \centering
    \includegraphics[width=1.\textwidth]{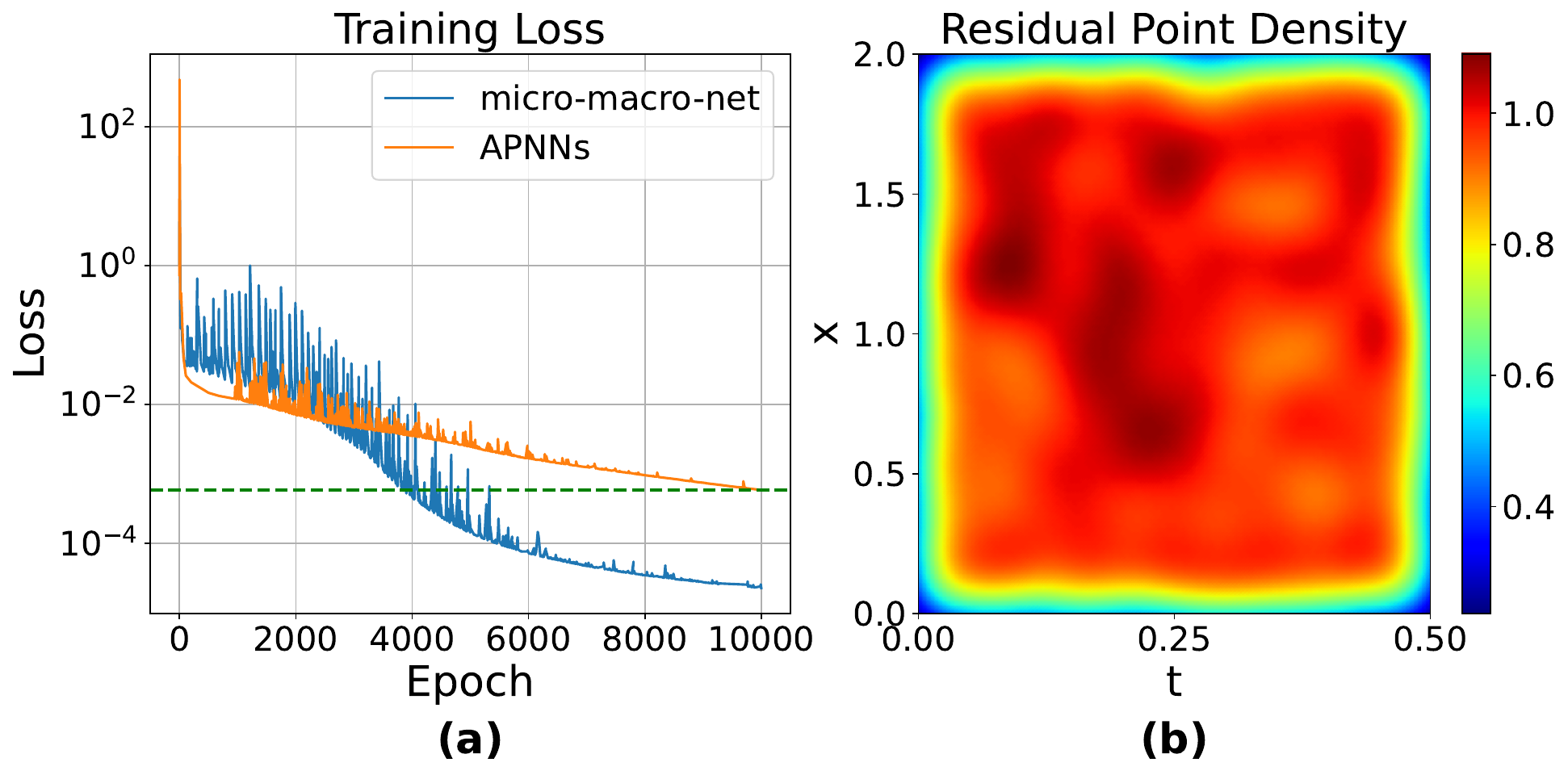}
    \caption{\@ The result of Ex 3 with $\epsilon$ = 1. 
    (a): The loss decay curves obtained by training with APNNs and APNNs+\ding{172}. (b): The distribution density function of the collocation points obtained by RT-APNN through MCMC adaptive sampling.}\label{fig:ex3_loss and density}
\end{figure}

First, we compare the loss descent curves of APNNs and micro-macro-net. 
As shown in Figure~\ref{fig:ex3_loss and density}(a), micro-macro-net significantly improves the convergence speed. 
Figure~\ref{fig:ex3_loss and density}(b) demonstrates that, due to the minimal variation in residuals across the solution domain, the MCMC method results in a relatively uniform distribution of collocation points.

\begin{figure}[]
    \centering
    \includegraphics[width=0.99\textwidth]{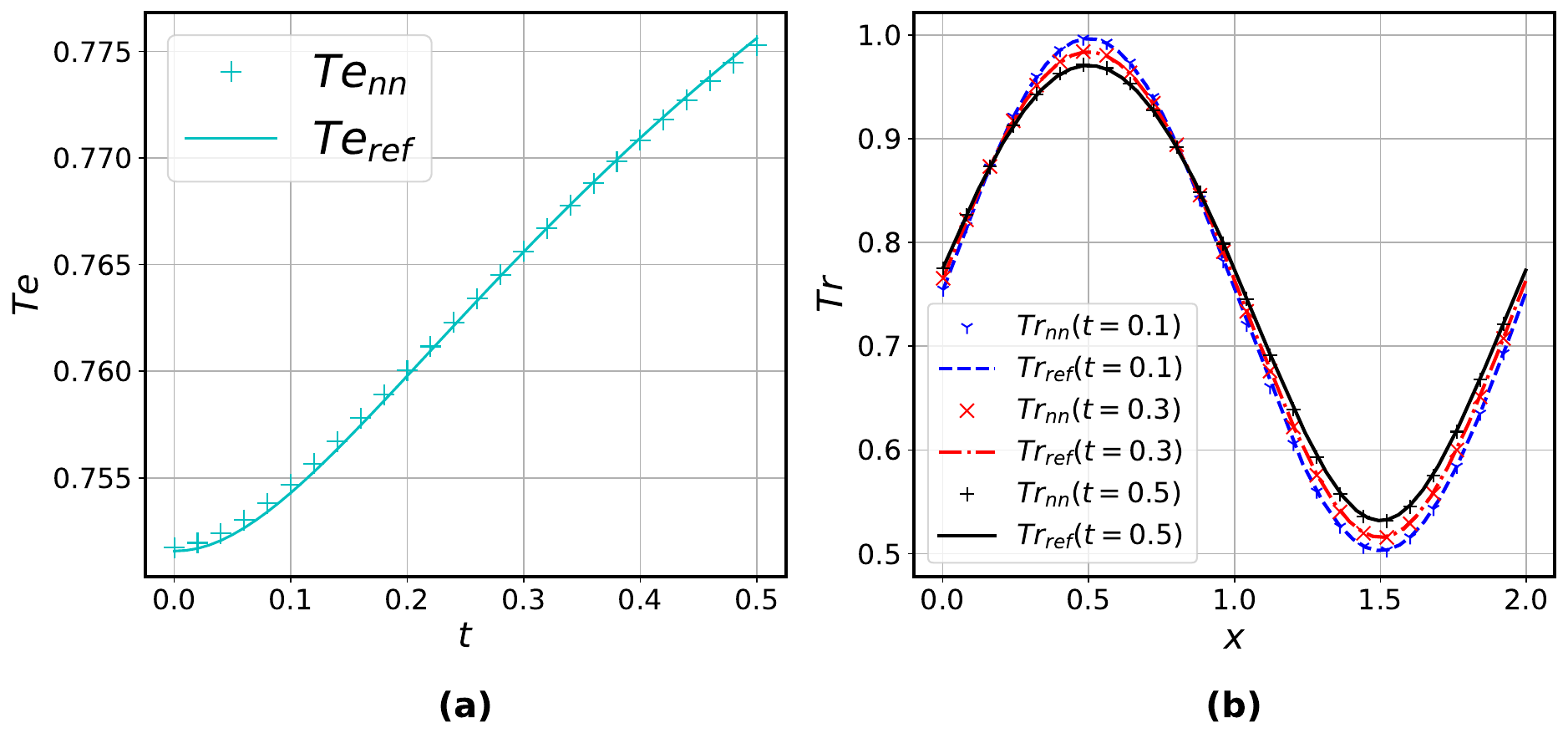}
    \caption{\@ The result of Ex 3. Comparison between the reference solution and RT-APNN when $\varepsilon=1$. (a): The material temperature $T_e$ at $x=0.0025$. (b): The radiation temperature $T_r$ at times $t=0.1,0.3,0.5$.}\label{fig:T_ex3}
\end{figure}

Figure~\ref{fig:T_ex3} compares the solutions obtained by RT-APNN with the reference solution.
It is clear that RT-APNN provides consistent results with the reference solution for both the radiation temperature \(T_r\) and the material temperature \(T_e\). 
Table~\ref{tab:error-ex3} presents the results for APNNs, MD-APNNs, and RT-APNN. Notably, RT-APNN achieves approximately an order of magnitude higher accuracy compared to MD-APNNs, which require additional supervised data points, despite not utilizing any extra supervision information.

\begin{figure}[]
    \centering
    \includegraphics[width=0.99\textwidth]{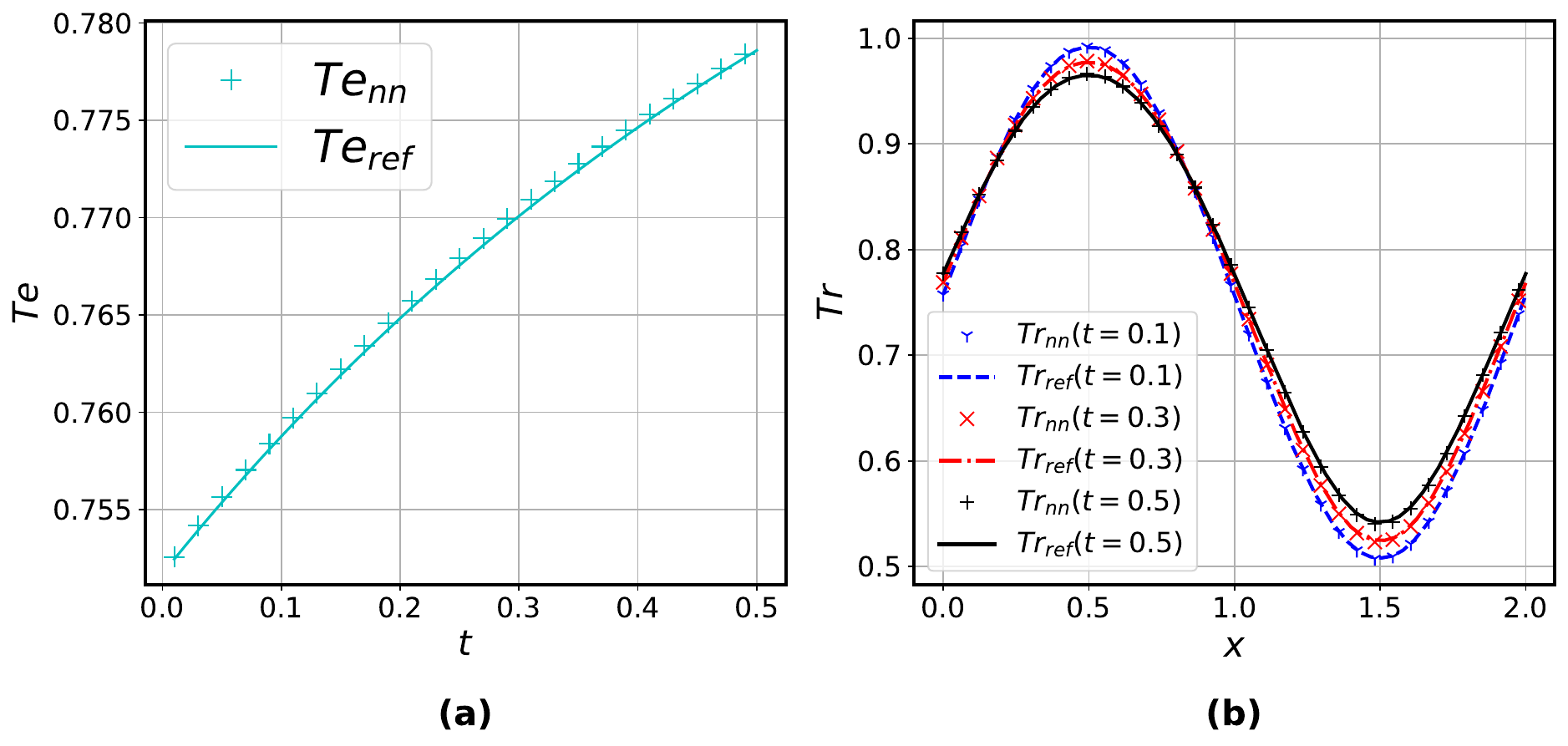}
    \caption{\@ The result of Ex 3. Comparison between the reference solution and RT-APNN when $\varepsilon=10^{-3}$. (a): The material temperature $T_e$ at $x=0.0025$. (b): The radiation temperature $T_r$ at times $t=0.1,0.3,0.5$.}\label{fig:T_1e-3_ex3}
\end{figure}

For the small-scale scenario with \(\varepsilon = 10^{-3}\), Figure~\ref{fig:T_1e-3_ex3} and Table~\ref{tab:error-1e-3-ex3} show that RT-APNN achieves significantly better accuracy compared to both APNNs and MD-APNNs.

\begin{small}
    \begin{table}[]
        \caption{Kinetic regime with $\varepsilon=1$ in Ex 3: Errors of $T_e$ (at $x= 0.0025$) and $T_r$ (at $t=0.1,0.2,0.3,0.4,0.5$) for APNNs, MD-APNNs, and RT-APNN  \cite{Li2022AMA}.}
        \label{tab:error-ex3}
        \centering
        \begin{tabular}{ccccccc}
            \toprule[1pt]$L^2$ error & $T_e(x=0.0025)$ & $T_r(t=0.1)$ & $T_r(t=0.2)$ & $T_r(t=0.3)$ & $T_r(t=0.4)$ & $T_r(t=0.5)$ \\
            \midrule 
            APNNs & $3.01 \mathrm{e}-03$ & $7.68 \mathrm{e}-03$ & $1.65 \mathrm{e}-02$ & $2.41 \mathrm{e}-02$ & $2.91 \mathrm{e}-02$ & $3.44 \mathrm{e}-02$ \\
            MD-APNNs & $4.24 \mathrm{e}-03$ & $3.12 \mathrm{e}-03$ & $3.62 \mathrm{e}-03$ & $3.92 \mathrm{e}-03$ & $4.80 \mathrm{e}-03$ & $6.79 \mathrm{e}-03$ \\
            RT-APNN & $3.25 \mathrm{e}-04$ & $3.78 \mathrm{e}-04$ & $5.78 \mathrm{e}-04$ & $8.18 \mathrm{e}-04$ & $9.99 \mathrm{e}-04$ & $1.20 \mathrm{e}-03$ \\
            \bottomrule
        \end{tabular}
    \end{table}
\end{small}

\begin{small}
    \begin{table}[]
        \caption{Intermediate regime with $\varepsilon=10^{-3}$ in Ex 3: Errors of $T_e$ (at $x= 0.0025$) and $T_r$ (at $t=0.1,0.2,0.3,0.4,0.5$) for APNNs, MD-APNNs, and RT-APNN  \cite{Li2022AMA}.}
        \label{tab:error-1e-3-ex3}
        \centering
        \begin{tabular}{ccccccc}
            \toprule[1pt]$L^2$ error & $T_e(x=0.0025)$ & $T_r(t=0.1)$ & $T_r(t=0.2)$ & $T_r(t=0.3)$ & $T_r(t=0.4)$ & $T_r(t=0.5)$ \\
            \midrule 
            APNNs & $2.82 \mathrm{e}-02$ & $7.10 \mathrm{e}-03$ & $1.16 \mathrm{e}-02$ & $1.54 \mathrm{e}-02$ & $1.89 \mathrm{e}-02$ & $2.23 \mathrm{e}-02$ \\
            MD-APNNs & $4.18 \mathrm{e}-03$ & $3.37 \mathrm{e}-03$ & $3.12 \mathrm{e}-03$ & $3.11 \mathrm{e}-03$ & $3.97 \mathrm{e}-03$ & $5.67 \mathrm{e}-03$ \\
            RT-APNN & $3.88 \mathrm{e}-04$ & $3.62 \mathrm{e}-04$ & $5.66 \mathrm{e}-04$ & $7.96 \mathrm{e}-04$ & $9.14 \mathrm{e}-04$ & $1.04 \mathrm{e}-03$ \\
            \bottomrule
        \end{tabular}
    \end{table}
\end{small}

\subsubsection{Ex 4: Nonlinear GRTEs with Reflective Boundary Conditions}

In this example, we solve a GRTE problem where the opacity is independent of temperature. The governing equation is expressed as:
\begin{equation}
    \label{eq: Ex4}
    \left\{\begin{aligned}
        &\frac{\varepsilon^2}{c} \frac{\partial I}{\partial t}+\varepsilon \mu \frac{\partial I}{\partial x}=\sigma\left(\frac{1}{2} a c T^4-I\right),(t, x, \mu) \in \mathbb{T} \times D \times[-1,1], \\
        &\varepsilon^2 C_v \frac{\partial T}{\partial t}=\sigma\left(2\langle I\rangle-a c T^4\right),(t, x) \in \mathbb{T} \times D, \\
        &I(t, 0, \mu>0)=I(t, 0,-\mu), \\
        &I(t, 0.25, \mu<0)=\frac{1}{2} a c(0.1)^4, \\
        &I(0, x, \mu)=\frac{1}{2} a c T(0, x)^4, \quad T(0, x)=1.
    \end{aligned}\right.
\end{equation}
where the opacity \(\sigma\) is \(10 \ \mathrm{cm}^{-1}\) and the heat capacity \(C_v\) is \(1 \ \mathrm{GJ/cm^3/keV}\). The problem is defined on a slab of length \(0.25 \ \mathrm{cm}\). Initially, the slab is in equilibrium at \(1 \ \mathrm{keV}\). The left boundary adopts a reflective condition, while the right boundary imposes an incident Planckian source condition. The radiation constant is \(a = 0.01372 \ \mathrm{GJ/cm^3/keV^4}\), and the speed of light is \(c = 29.98 \ \mathrm{cm/ns}\).

In \eqref{eq: Ex4}, the radiation intensity at the right boundary at \(t = 0 \ \mathrm{ns}\) differs from the radiation intensity in the initial condition at \(x = 0.25 \ \mathrm{cm}\). This represents a typical initial-boundary value incompatibility problem, which poses significant challenges for solving equations using neural networks. The mismatch between initial and boundary conditions can result in numerical instability or convergence failure during training. 

\begin{figure}[]
    \centering
    \includegraphics[width=0.99\textwidth]{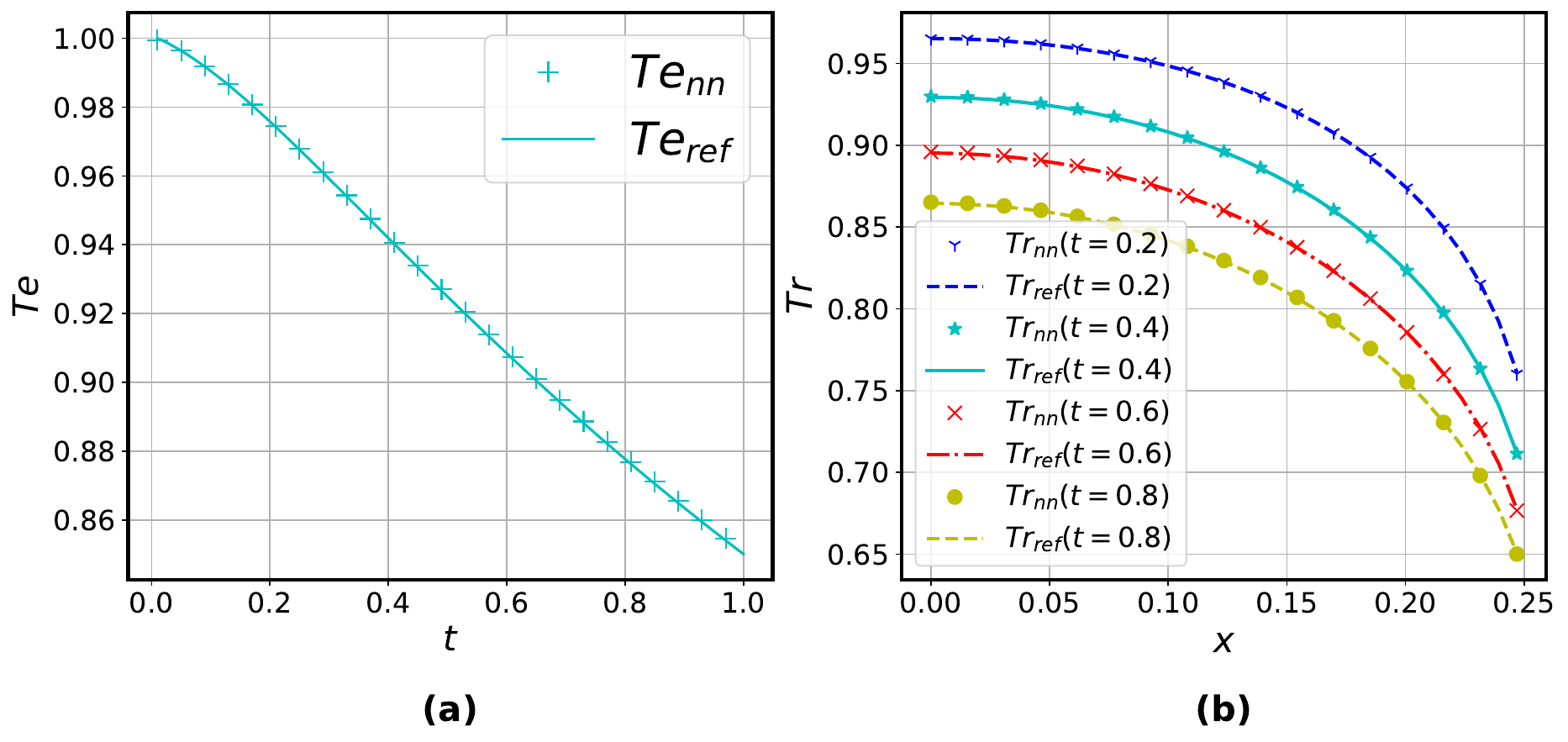}
    \caption{\@ The result of Ex 4. Comparison between RT-APNN and reference solutions (lines) with $\varepsilon=1$. (a): The material temperature at $x=0.0025$. (b): The radiation temperatures at times $t=0.2, 0.4, 0.6, 0.8$.}\label{fig:T_ex4}
\end{figure}

As shown in Figure~\ref{fig:T_ex4}, the proposed network architecture and loss function enable accurate solutions even in the presence of such incompatibilities. Additionally, Table~\ref{tab:error-ex4} highlights that RT-APNN achieves excellent predictive accuracy. These results not only demonstrate the effectiveness of RT-APNN but also address the unresolved issue identified in \cite{Li2022AMA}, namely, eliminating the dependence on additional supervised data.

\begin{small}
    \begin{table}[]
        \caption{Kinetic regime with $\varepsilon=1$ in Ex 4: Errors of $T_e$ (at $x= 0.0025$) and $T_r$ (at $t=0.2, 0.4, 0.6, 0.8$) for RT-APNN.}
        \label{tab:error-ex4}
        \centering
        \begin{tabular}{cccccc}
            \toprule$L^2$ error & $T_e(x=0.0025)$ & $T_r(t=0.2)$ & $T_r(t=0.4)$ & $T_r(t=0.6)$ & $T_r(t=0.8)$ \\
            \midrule RT-APNN & $4.72 \mathrm{e}-04$ & $4.14 \mathrm{e}-04$ & $4.77 \mathrm{e}-04$ & $5.92 \mathrm{e}-04$ & $6.10 \mathrm{e}-04$
 \\
            \bottomrule
        \end{tabular}
    \end{table}
\end{small}

\subsubsection{Ex 5: The Marshak Wave problem}
The Marshak Wave problem is a classical test case in radiation hydrodynamics, examining the propagation of a radiation wave through a medium with specified material properties and boundary conditions. It serves as a benchmark for evaluating the accuracy and performance of numerical methods and codes used in solving radiation transport equations, with applications in fields such as astrophysics, nuclear engineering, and inertial confinement fusion. 

The governing equations for the Marshak Wave problem are given by:
\begin{equation}
    \left\{
    \begin{aligned}
        &\frac{\varepsilon^2}{c} \frac{\partial I}{\partial t}+\varepsilon \mu \frac{\partial I}{\partial x}=\sigma\left(\frac{1}{2} a c T^4-I\right),(t, x, \mu) \in \mathbb{T} \times D \times[-1,1], \\
        &\varepsilon^2 C_v \frac{\partial T}{\partial t}=\sigma\left(2\langle I\rangle-a c T^4\right),(t, x) \in \mathbb{T} \times D, \\
        &I\left(t, x_L, \mu>0\right)=\frac{1}{2} a c T_{b d}^4, \\
        &I\left(t, x_R, \mu<0\right)=\frac{1}{2} a c T_{0}^4, \\
        &I(0, x, \mu)=\frac{1}{2} a c T_0^4.
    \end{aligned}
    \right.
\end{equation}
To simulate the problem, we use \( c = 29.98 \ \mathrm{cm/ns} \), \( a = 0.01372 \ \mathrm{GJ/cm^3/keV^4} \), and \( C_v = 0.3 \ \mathrm{GJ/cm^3/keV} \). The absorption coefficient is given by \( \sigma_a(T) = \frac{\sigma_{a, 0}}{\left( T / T_{\mathrm{keV}} \right)^3} \), where \( \sigma_{a, 0} = 30 \ \mathrm{cm^{-1}} \), and \( T_{\mathrm{keV}} \) is chosen such that \( k_B T_{\mathrm{keV}} = 1 \ \mathrm{keV} \), with \( k_B \) being the Boltzmann constant. The initial condition is set to \( {T_0}/{T_{\mathrm{keV}}} = 10^{-2} \). The computational domain is \( [0 \ \mathrm{cm}, \infty \ \mathrm{cm}) \), but for the simulation, it is restricted to \( [0 \ \mathrm{cm}, 0.5 \ \mathrm{cm}] \), with inflow boundary conditions imposed at both ends, where \( T_{\mathrm{bd}}/{T_{\mathrm{keV}}} = 1 \). The problem is solved over the time interval \( \mathbb{T} = [0 \ \mathrm{ns}, 1 \ \mathrm{ns}] \).

\begin{figure}[]
    \centering
    \includegraphics[width=0.99\textwidth]{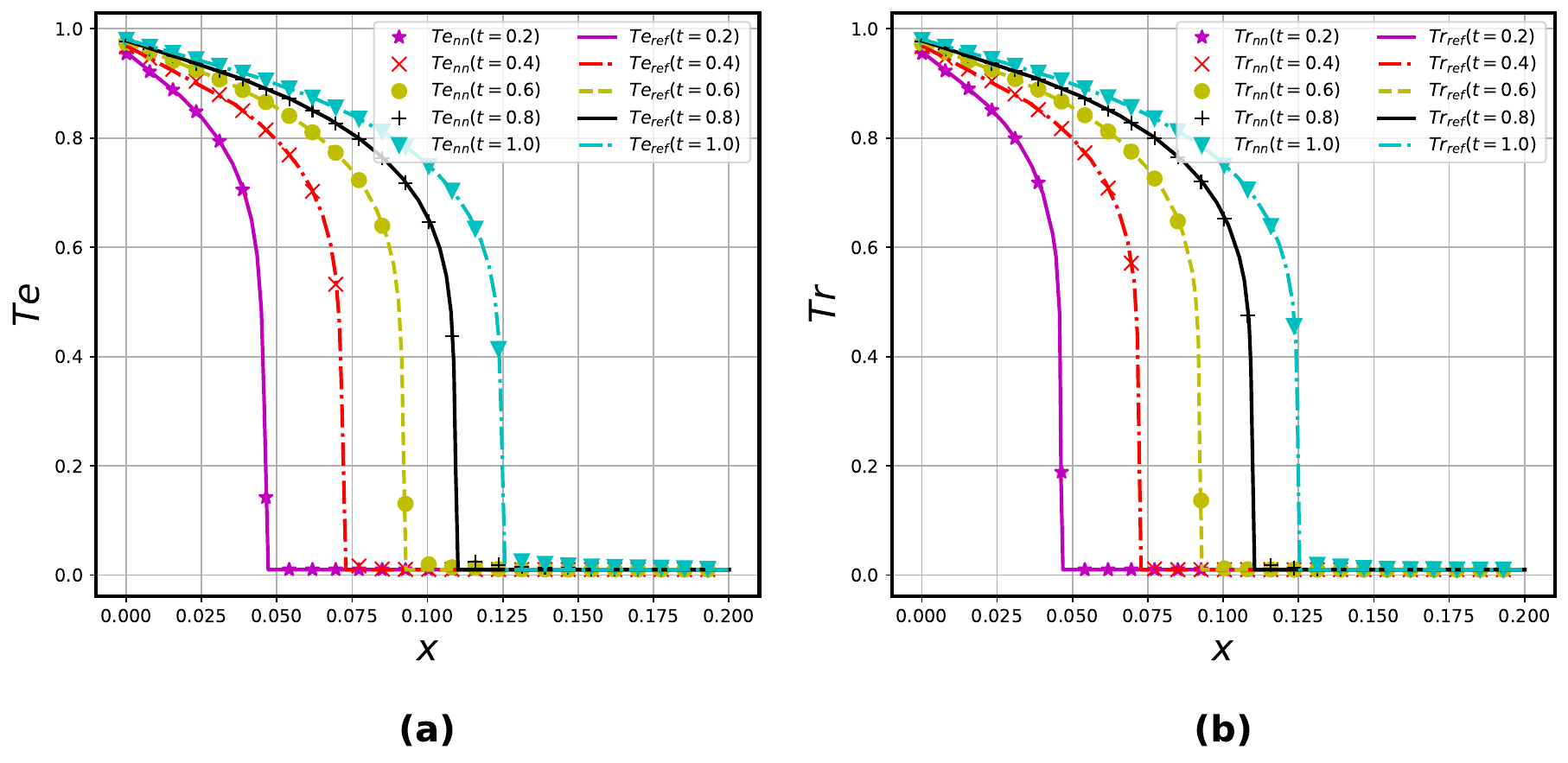}
    \caption{\@ The result of Ex 5. Comparison between RT-APNN and reference solutions (lines) with $\sigma_{a,0}=30\ \mathrm{cm^{-1}}$. (a): The material temperature. (b): The radiation temperatures at times $t=0.2\ \mathrm{ns}, 0.4\ \mathrm{ns}, 0.6\ \mathrm{ns}, 0.8\ \mathrm{ns},1.0\ \mathrm{ns}$.}\label{fig:T_ex5_30}
\end{figure}

Due to the intense energy exchange at the initial moment, leading to a sharp rise in material temperature and causing singularities in the problem, the simulation begins at \( t = 0.1 \ \mathrm{ns} \). The reference solution in Figure~\ref{fig:T_ex5_30} is obtained using the spherical harmonic Pn method. The results from the RT-APNN method are consistent with this deterministic approach and successfully capture the evolution near the wavefront.

As shown in Table~\ref{tab:error-ex5}, the relative errors in both radiation temperature and material temperature are approximately $1\%$, which is a satisfactory result. This is attributed to the large gradient near the wavefront, where even a slight deviation in the wavefront position can lead to a significant increase in the relative error.

\begin{small}
    \begin{table}[]
        \caption{Kinetic regime with $\sigma_{a,0}=30 \ \mathrm{cm^{-1}} $ in Ex 5: Errors of $T_e$ and $T_r$ (at $t=0.2\ \mathrm{ns}, 0.4\ \mathrm{ns}, 0.6\ \mathrm{ns}, 0.8\ \mathrm{ns},1.0\ \mathrm{ns}$) for RT-APNN.}
        \label{tab:error-ex5}
        \centering
        \begin{center}
            \begin{tabular}{cccccc}
                \toprule$L^2$ error & $t=0.2$ & $t=0.4$ & $t=0.6$ & $t=0.8$& $t=1.0$ \\
                \midrule 
                Te & $1.75 \mathrm{e}-02$ & $1.74 \mathrm{e}-02$ & $1.82 \mathrm{e}-02$ & $2.01 \mathrm{e}-02$ & $1.85 \mathrm{e}-02$ \\
                Tr & $1.76 \mathrm{e}-02$ & $1.54 \mathrm{e}-02$ & $1.65 \mathrm{e}-02$ & $1.83 \mathrm{e}-02$ & $1.73 \mathrm{e}-02$ \\
                \bottomrule
            \end{tabular}
        \end{center}

    \end{table}
\end{small}

\begin{figure}[]
    \centering
    \includegraphics[width=0.9\textwidth]{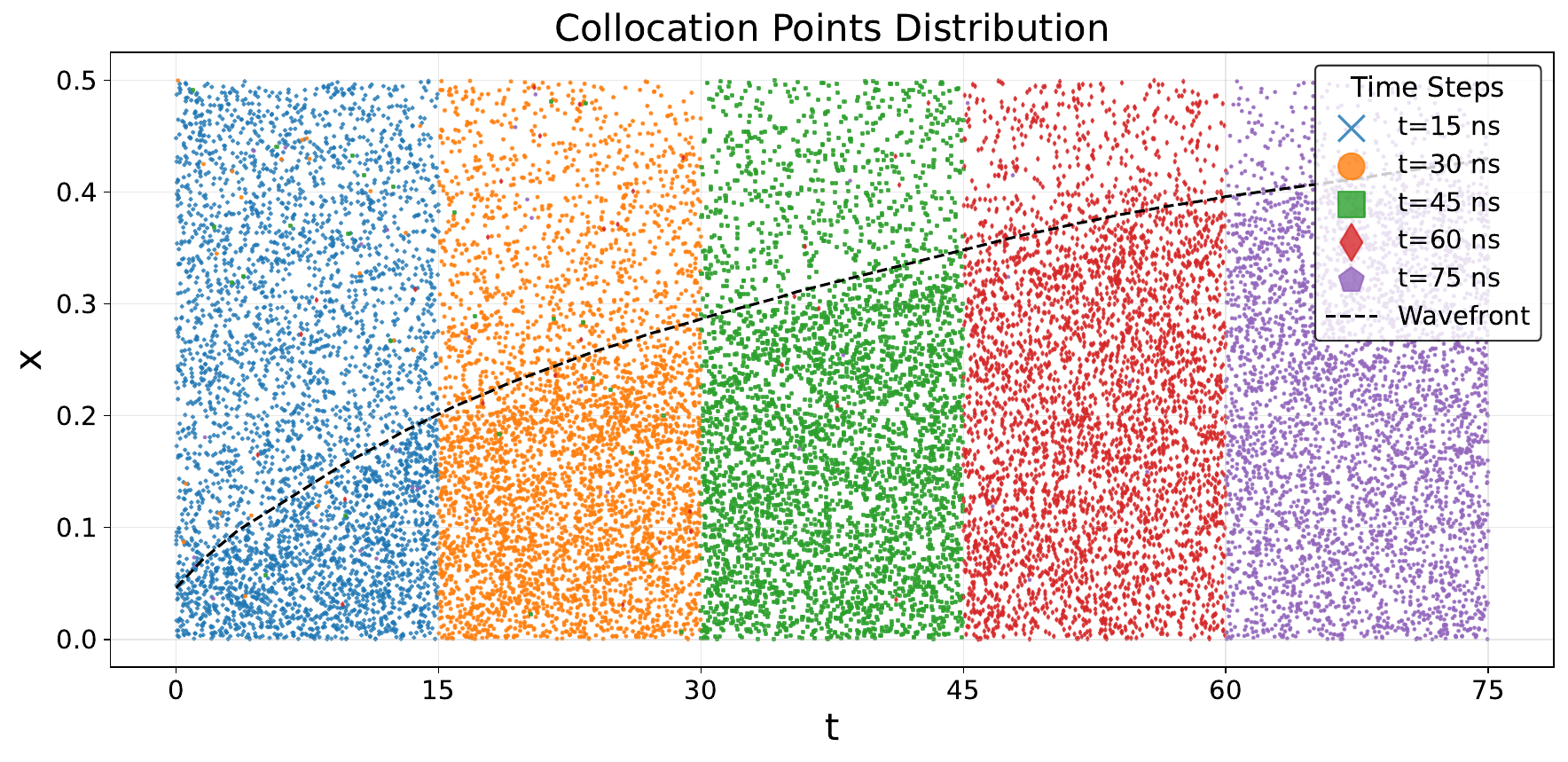}
    \caption{\@ Scatter plot of collocation points with $\sigma_{a,0}=300 \ \mathrm{cm^{-1}} $ in Ex 5: Using a pre-training step of 15$\ \mathrm{ns}$ as an example, The positions of collocation points under the influence of MCMC at the pre-training strategy times of 15$\ \mathrm{ns}$, 30$\ \mathrm{ns}$, 45$\ \mathrm{ns}$, 60$\ \mathrm{ns}$, and 75$\ \mathrm{ns}$, respectively.}\label{fig:point_ex5_300}
\end{figure}

Finally, we tackle a more challenging experiment by setting \( \sigma_{a, 0} = 300 \ \mathrm{cm}^{-1} \) and solving over the time domain \( \mathbb{T} = [0 \ \mathrm{ns}, 75 \ \mathrm{ns}] \). Our pre-training strategy employs a time step of \( 15 \ \mathrm{ns} \), with \( 90\% \) of the sampled collocation points concentrated in untrained regions during each pre-training phase. 
The distribution of collocation points is illustrated in detail in Figure~\ref{fig:point_ex5_300}. 
The adaptive sampling method based on MCMC ensures that these collocation points are precisely located in the region around the wavefront, achieving effective local refinement. 
It also reduces the number of sampling points in regions where the wave has not yet reached, allowing the problem to be solved with a minimal number of collocation points, significantly improving computational efficiency.

\begin{figure}[]
    \centering
    \includegraphics[width=0.99\textwidth]{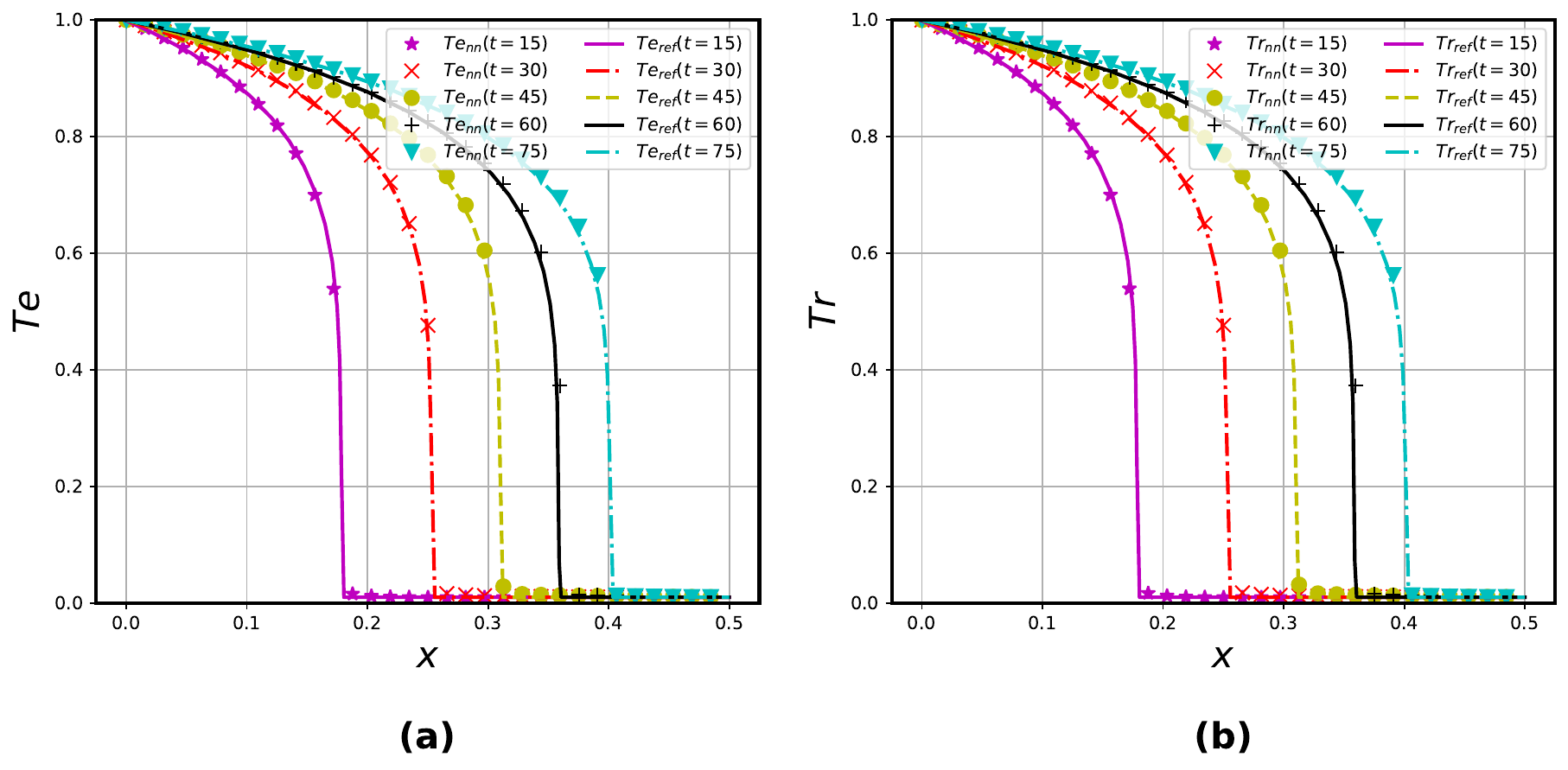}
    \caption{\@ The result of Ex 5. Comparison between RT-APNN and reference solutions (lines) with $\sigma_{a,0}=300\ \mathrm{cm^{-1}}$. (a): The material temperature, (b): The radiation temperatures at times $t=1\ \mathrm{ns}, 15\ \mathrm{ns}, 30\ \mathrm{ns}, 45\ \mathrm{ns},60\ \mathrm{ns},75\ \mathrm{ns}$.}\label{fig:T_ex5_300}
\end{figure}

The results are presented in Figure~\ref{fig:T_ex5_300} and Table~\ref{tab:error-ex5_300}. Notably, the dependence of the absorption coefficient on material temperature, an issue unresolved in \cite{Li2022AMA}, is successfully addressed in this work. RT-APNN is the first machine learning method capable of solving this problem.

In this experiment, both the MCMC method and the pre-training strategy are crucial. The computational domain in this case is 75 times larger in time compared to the scenario where \( \sigma_{a, 0} = 30 \ \mathrm{cm}^{-1} \), and 2.5 times larger in space. Consequently, the density of collocation points reduces to \( \frac{1}{187.5} \) while maintaining the same total number of points. The MCMC method enhances the efficiency of the collocation points, while the pre-training strategy enables decomposition of the problem into smaller scales, making it possible to address long time-domain simulations effectively.

\begin{small}
    \begin{table}[]
        \caption{Kinetic regime with $\sigma_{a,0}=300 \ \mathrm{cm^{-1}} $ in Ex 5: Errors of $T_e$ and $T_r$ (at $t=15\ \mathrm{ns}, 30\ \mathrm{ns}, 45\ \mathrm{ns}, 60\ \mathrm{ns},75\ \mathrm{ns}$) for RT-APNN.}
        \label{tab:error-ex5_300}
        \centering
        \begin{center}
            \begin{tabular}{cccccc}
                \toprule$L^2$ error & $t=15$ & $t=30$ & $t=45$ & $t=60$& $t=75$ \\
                \midrule 
                Te  & $4.39 \mathrm{e}-02$ & $6.18 \mathrm{e}-03$ & $2.71 \mathrm{e}-02$ & $3.22 \mathrm{e}-02$ & $2.93 \mathrm{e}-02$ \\
                Tr  & $4.29 \mathrm{e}-02$ & $6.80 \mathrm{e}-03$ & $2.76 \mathrm{e}-02$ & $3.23 \mathrm{e}-02$ & $2.95 \mathrm{e}-02$ \\
                \bottomrule
            \end{tabular}
        \end{center}

    \end{table}
\end{small}

\subsection{Solve the 2D radiation transport equation}

\subsubsection{Ex 6: Test Case with Smooth Initial Conditions and Periodic Boundary Conditions}\label{sec:ex6}
Previous studies have primarily focused on solving the one-dimensional radiative transfer equation. In this work, we extend the evaluation to a two-dimensional nonlinear scenario, where the governing system of equations is defined as follows:

\begin{equation}
\left\{
\begin{aligned}
&\frac{\partial \rho}{\partial t} + \frac{c}{\sqrt{\sigma_0}} \nabla \cdot \langle \Omega g \rangle = \frac{c \tilde{\sigma}}{\varepsilon^2 / \sigma_0} (\operatorname{acT}^4 /|\Omega| - \rho), \\
&\frac{\varepsilon^2}{\sigma_0} C_v \frac{\partial T}{\partial t} = \tilde{\sigma} |\Omega| (\rho - \operatorname{acT}^4 /|\Omega|), \\
&g_t + \frac{c}{\varepsilon} \left(\nabla \cdot (\Omega g)+\nabla \cdot \langle \Omega g \rangle\right) + \frac{c \sqrt{\sigma_0}}{\varepsilon^2} \nabla \cdot (\Omega \rho) = -\frac{c \tilde{\sigma}}{\varepsilon^2 / \sigma_0} g, \\
&\rho(0, x, y) = (a_1 + b_1 \sin(x))(a_2 + b_2 \sin(y))^4, \\
&T(0, x, y) = (a_1 + b_1 \sin(x))(a_2 + b_2 \sin(y)), \\
&g(0, x, y) = -\Omega \cdot \nabla \rho(0, x, y) / \sigma,
\end{aligned}
\right.
\end{equation}
where \(\tilde{\sigma} = \sigma / \sigma_0\), with \(\sigma = \sigma_0 = 1\), and the constant parameters \(a\), \(c\), and \(C_v\) are all set to 1. Specifically, \(a_1 = a_2 = 0.8\) and \(b_1 = b_2 = 0.1\). The computation is performed over the time domain \(\mathbb{T} = [0, 1]\) and the spatial domain \([0, 2\pi]\), with periodic boundary conditions applied in the spatial directions.

\begin{figure}[]
    \centering
    \includegraphics[width=0.98\textwidth]{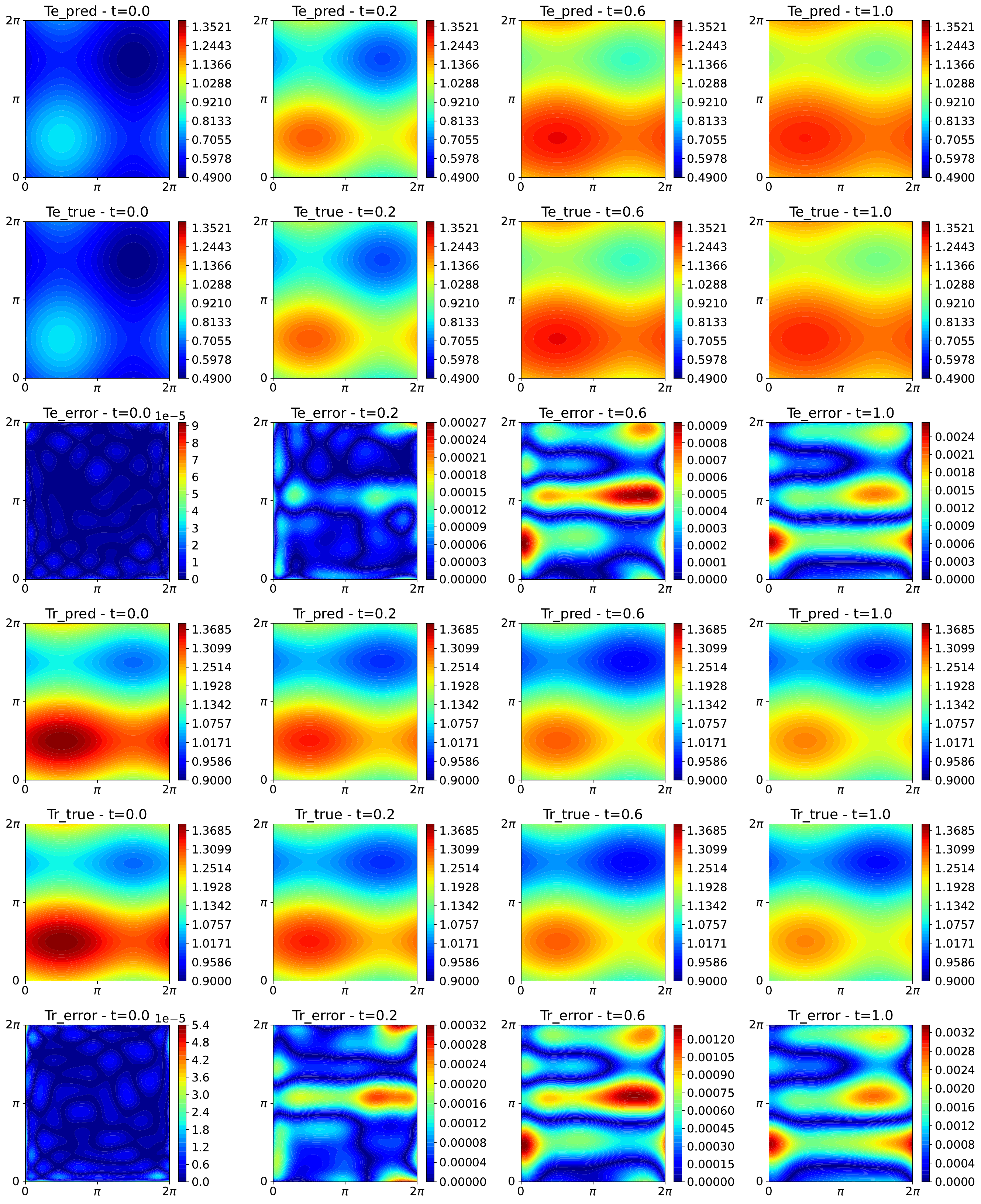}
    \caption{\@ Plot of pointwise errors between RT-APNN and the reference solution in Ex 6. Time progresses from left to right: $t=0, 0.2, 0.6, 1.0$.}\label{fig:error_ex6}
\end{figure}

In this example, the integration method employs 50 Lebedev quadrature points. During the integration, elements of the rotation group \(SO(3)\) act on these points, causing the integration points to dynamically adjust while maintaining their relative positions unchanged. The solution results, shown in Figure~\ref{fig:error_ex6} and Table~\ref{tab:error-ex6}, confirm that RT-APNN effectively handle smooth, high-dimensional radiative transport problems.

\begin{small}
\begin{table}[]
    \caption{Kinetic regime with $\varepsilon=1$ in Ex 6: Errors of $T_e$ and $T_r$ for RT-APNN solving the 2D problem at $t=0, 0.2, 0.4, 0.6, 0.8, 1.0$.}
    \label{tab:error-ex6}
    \centering
    \begin{center}
            \begin{tabular}{ccccccc}
                \toprule$L^2$ error & $t=0$ & $t=0.2$ & $t=0.4$ & $t=0.6$ & $t=0.8$ & $t=1.0$ \\
                \midrule 
                $T_e$ & $7.52 \mathrm{e}-06$ & $5.09 \mathrm{e}-05$ & $1.65 \mathrm{e}-04$ & $3.45 \mathrm{e}-04$ & $5.88 \mathrm{e}-04$ & $9.11 \mathrm{e}-04$ \\
                $T_r$ & $4.24 \mathrm{e}-06$ & $8.84 \mathrm{e}-05$ & $2.62 \mathrm{e}-04$ & $5.03 \mathrm{e}-04$ & $8.14 \mathrm{e}-04$ & $1.20 \mathrm{e}-03$ \\
                \bottomrule
            \end{tabular}
    \end{center}
\end{table}
\end{small}

\section{Conclusion}
Gray Radiative Transfer Equations (GRTEs) are fundamental in numerous scientific and engineering applications involving complex thermal radiation processes. 
This paper introduces the Radiative Transfer Asymptotically Preserving Neural Network (RT-APNN), a novel framework specifically designed to solve GRTEs.
By enhancing the network architecture, RT-APNN simplifies the training process, reduces computational costs, and improves efficiency, all while maintaining or enhancing solution accuracy.

For solving the Marshak wave problem, which involves a large temperature gradient, pre-training techniques are employed to enhance the robustness of long-term problem-solving. 
Combined with Markov Chain Monte Carlo (MCMC) adaptive sampling, which optimizes the distribution of collocation points based on problem characteristics, the method successfully captures the wavefront position.
This marks the first machine learning approach to successfully solve the Marshak wave problem. 
These innovations also demonstrate exceptional efficiency in solving other time-dependent nonlinear GRTEs and high-dimensional problems.
Numerical experiments reveal that RT-APNN outperforms existing methods, including APNNs and MD-APNNs, in both accuracy and computational efficiency across multiple test cases.

However, this study considers temperature-dependent media, namely Marshak Wave. 
We still cannot solve the Marshak Wave from the initial moment because neural networks cannot represent functions with incompatible initial-boundary values. 
Moreover, for high-dimensional temperature-dependent media problems, although the micro-macro network improves convergence speed, the solution still requires numerous internal configuration points and spherical surface integration points.
Furthermore, with updates to network parameters, spherical numerical integration must be calculated at each internal collocation point, posing challenges in memory requirements and computational efficiency.
Reducing these computational overheads and addressing high-dimensional complex problem will be key areas of future research.

In summary, RT-APNN offers a powerful framework, with its techniques and strategies potentially applicable to other types of PDEs.
This research not only advances methods for solving GRTEs but also presents a practical approach that harnesses machine learning to address complex physical processes, highlighting the significant potential of machine learning-based simulations for GRTEs.

\begin{appendices} 
\section{The number of quadrature points and collocation points affects the results}\label{app:a} 
We conduct tests on three sets of experiments, \ref{sec:ex1} \ref{sec:ex3}, and \ref{sec:ex6}. Different numbers of quadrature nodes and boundary points are selected (with the number of internal configuration points being four times that of boundary points). 
The geometric mean of the relative error of temperature at different moments is ultimately presented.
From Figures~\ref{fig:Ex 1 Tr Error (integral_nodes,num_points)} and \ref{fig:Ex 3 Tr Error (integral_nodes,num_points)}, it is evident that for one-dimensional problems, using only one Gaussian-Legendre quadrature point leads to solver failure, while using three or more quadrature points does not significantly improve the solution's accuracy. From Figures \ref{fig:Ex 6 Te Error (integral_nodes,num_points)} and \ref{fig:Ex 6 Tr Error (integral_nodes,num_points)}, it is clear that for high-dimensional problems, the number of sampling points has a significant impact on solution accuracy. In cases with fewer sampling points, increasing the number of quadrature points does not improve accuracy. However, when the number of sampling points is large, even using only six Lebedev quadrature points can yield good results, and accuracy can be further improved by increasing the number of quadrature points.

\begin{figure}[]
    \centering
    \includegraphics[width=0.98\textwidth]{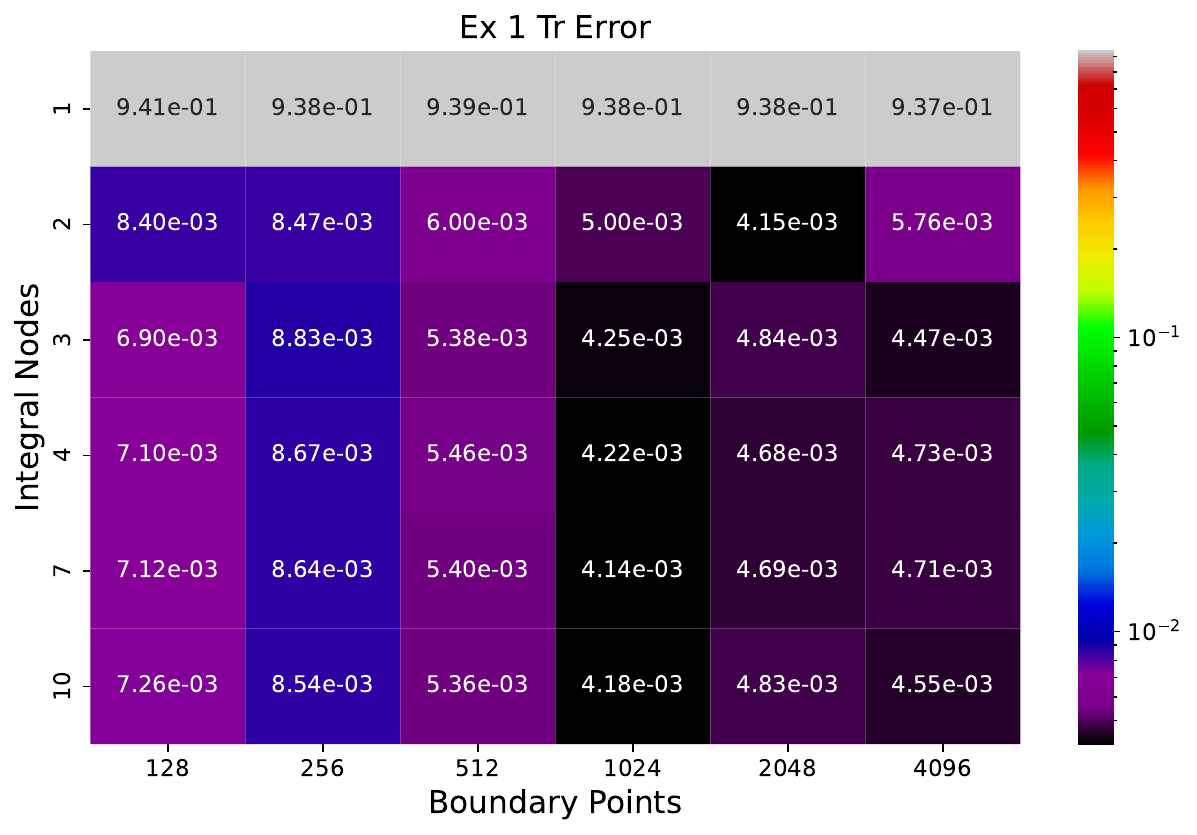}
    \caption{\@ $T_r$ errors with different numbers of Gaussian-Legendre quadrature points and sampling points for Ex 1.}\label{fig:Ex 1 Tr Error (integral_nodes,num_points)}
\end{figure}

\begin{figure}[]
    \centering
    \includegraphics[width=0.98\textwidth]{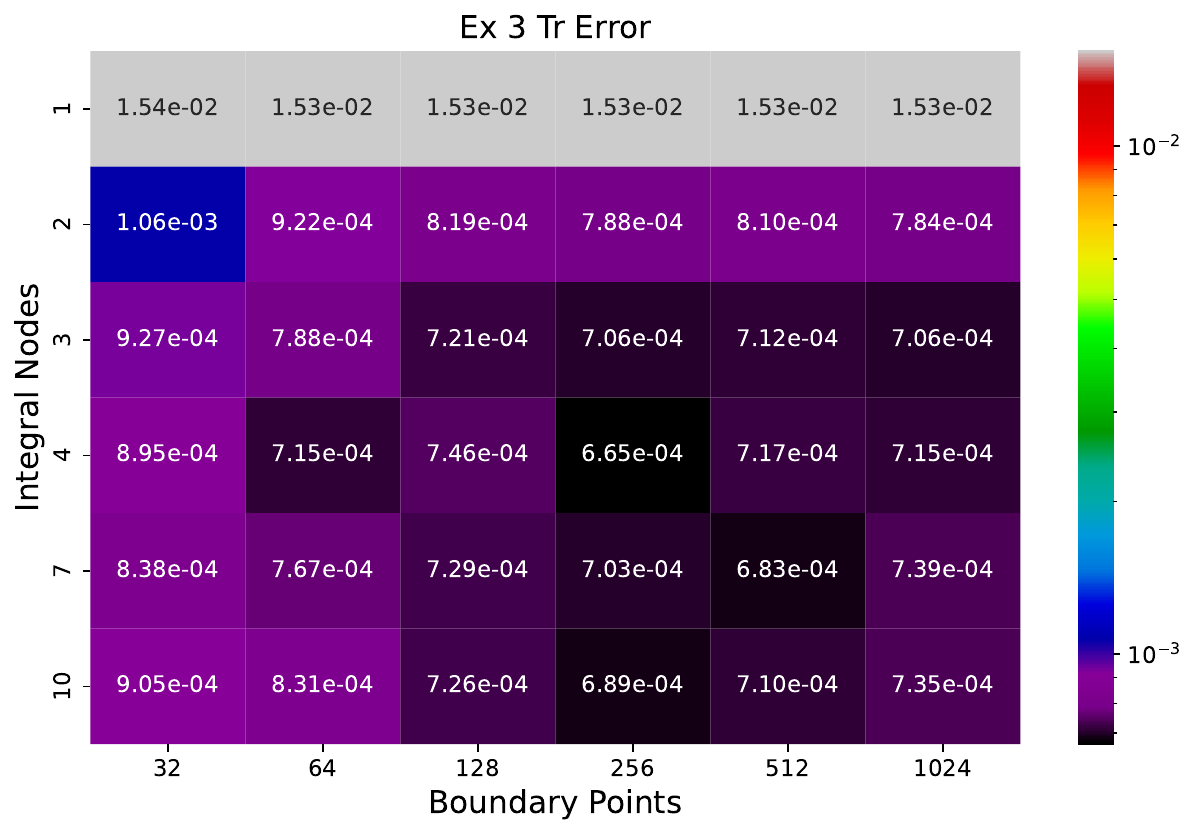}
    \caption{\@ $T_r$ errors with different numbers of Gaussian-Legendre quadrature points and sampling points for Ex 3.}\label{fig:Ex 3 Tr Error (integral_nodes,num_points)}
\end{figure}

\begin{figure}[]
    \centering
    \includegraphics[width=0.98\textwidth]{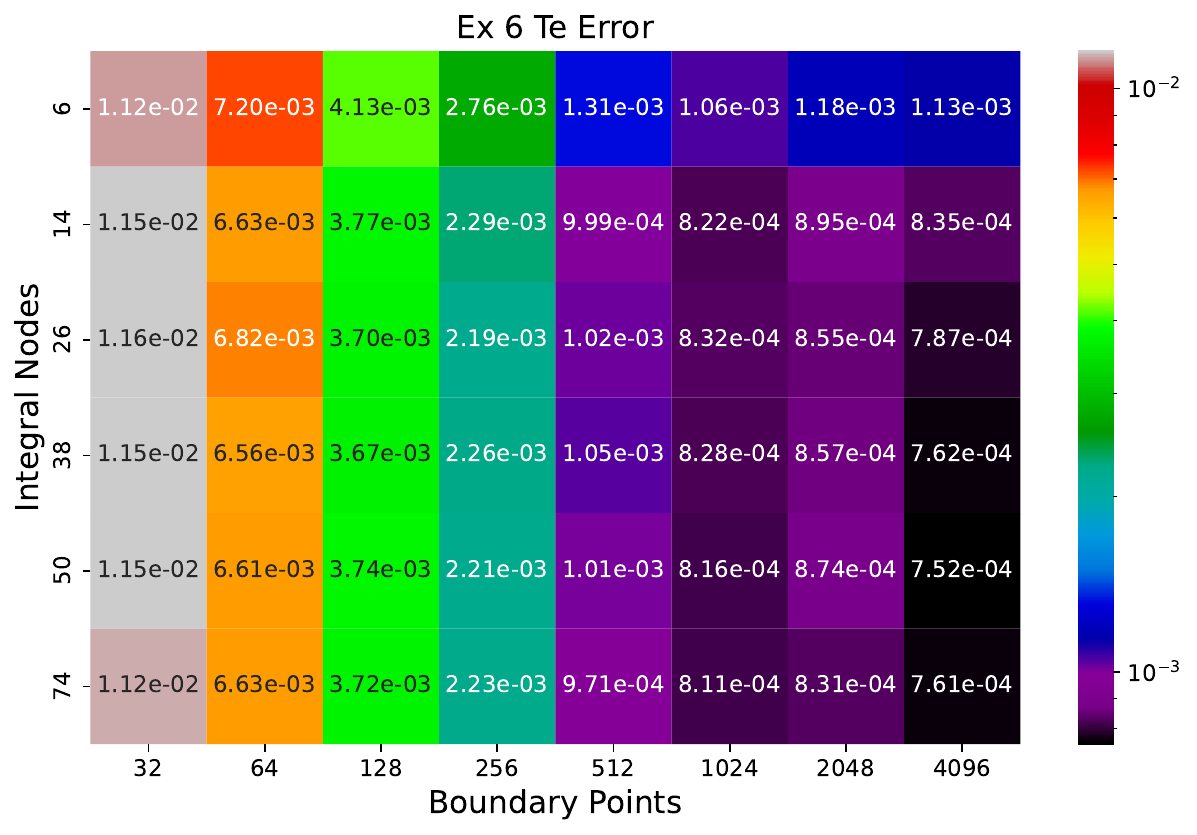}
    \caption{\@ $T_e$ errors with different numbers of Lebedev quadrature points and sampling points for Ex 6.}\label{fig:Ex 6 Te Error (integral_nodes,num_points)}
\end{figure}

\begin{figure}[]
    \centering
    \includegraphics[width=0.98\textwidth]{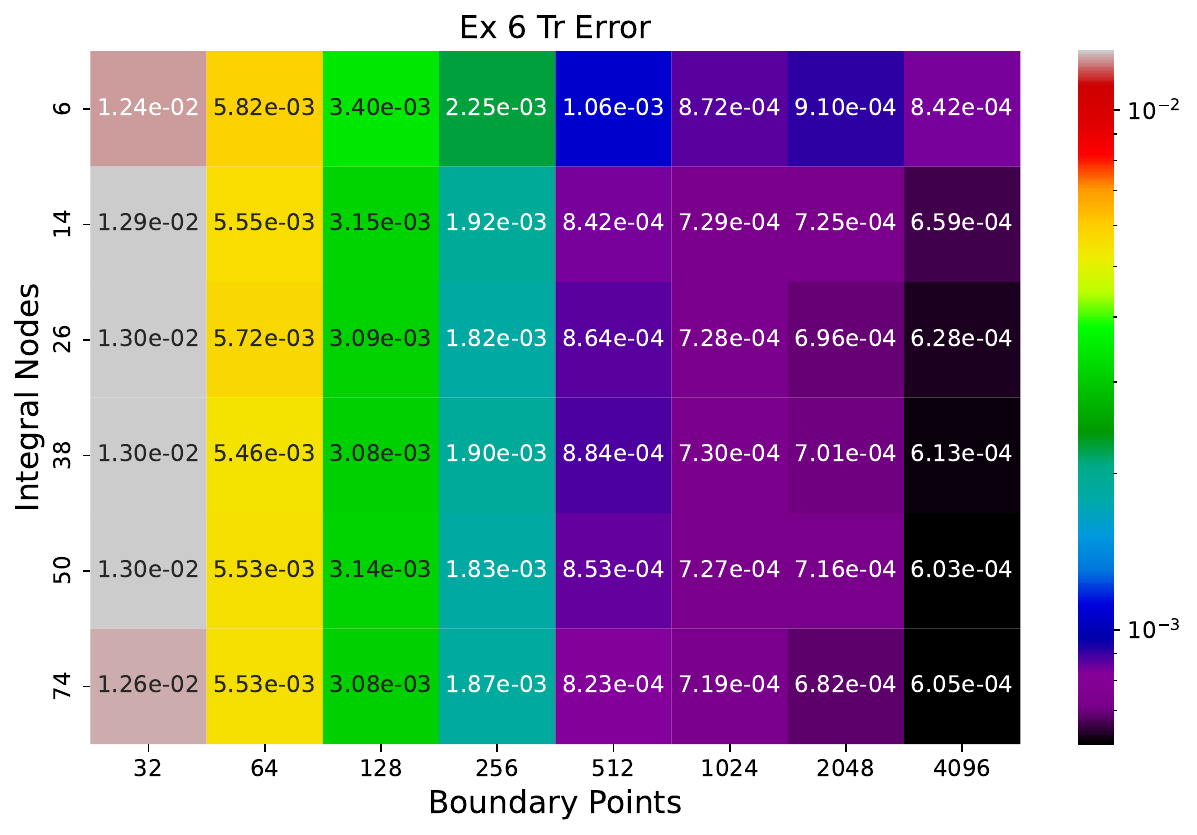}
    \caption{\@ $T_r$ errors with different numbers of Lebedev quadrature points and sampling points for Ex 6.}\label{fig:Ex 6 Tr Error (integral_nodes,num_points)}
\end{figure}

\end{appendices}

\section{Declaration of competing interest}
The authors declare that they have no known competing financial interests or personal relationships that could have appeared to influence the work reported in this paper.

\section{Data availability}
My manuscript has no associated data.

\bibliographystyle{plain}
\bibliography{ref.bib}

\begin{thebibliography}{10}

\bibitem{carlson1955solution}
Bengt~G Carlson.
\newblock Solution of the transport equation by sn approximations.
\newblock Technical report, Los Alamos National Lab.(LANL), Los Alamos, NM
  (United States), 1955.

\bibitem{chandrasekhar2013radiative}
Subrahmanyan Chandrasekhar.
\newblock {\em Radiative transfer}.
\newblock Courier Corporation, 2013.

\bibitem{cheng2021deep}
Chen Cheng and Guang-Tao Zhang.
\newblock Deep learning method based on physics informed neural network with
  resnet block for solving fluid flow problems.
\newblock {\em Water}, 13(4):423, 2021.

\bibitem{clough2005atmospheric}
Shepard~A Clough, Mark~W Shephard, Eli~J Mlawer, JS~Delamere, Michael~J Iacono,
  Karen Cady-Pereira, S~Boukabara, and Patrick~D Brown.
\newblock Atmospheric radiative transfer modeling: A summary of the aer codes.
\newblock {\em Journal of Quantitative Spectroscopy and Radiative Transfer},
  91(2):233--244, 2005.

\bibitem{dahl1999positive}
JA~Dahl, BD~Ganapol, and JE~Morel.
\newblock Positive scattering cross sections using constrained least squares.
\newblock Technical report, Los Alamos National Lab.(LANL), Los Alamos, NM
  (United States), 1999.

\bibitem{DENSMORE20111116}
Jeffery~D. Densmore.
\newblock Asymptotic analysis of the spatial discretization of radiation
  absorption and re-emission in implicit monte carlo.
\newblock {\em Journal of Computational Physics}, 230(4):1116--1133, 2011.

\bibitem{densmore2004asymptotic}
Jeffery~D. Densmore and Edward~W Larsen.
\newblock Asymptotic equilibrium diffusion analysis of time-dependent monte
  carlo methods for grey radiative transfer.
\newblock {\em Journal of Computational Physics}, 199(1):175--204, 2004.

\bibitem{fleck1961calculation}
Joseph~A Fleck~Jr.
\newblock The calculation of nonlinear radiation transport by a monte carlo
  method.
\newblock Technical report, Lawrence Livermore National Lab.(LLNL), Livermore,
  CA (United States), 1961.

\bibitem{FLECK1971313}
Joseph~A Fleck~Jr and JD~Cummings~Jr.
\newblock An implicit monte carlo scheme for calculating time and frequency
  dependent nonlinear radiation transport.
\newblock {\em Journal of Computational Physics}, 8(3):313--342, 1971.

\bibitem{FRANK20072289}
Martin Frank, Axel Klar, Edward~W. Larsen, and Shugo Yasuda.
\newblock Time-dependent simplified pn approximation to the equations of
  radiative transfer.
\newblock {\em Journal of Computational Physics}, 226(2):2289--2305, 2007.

\bibitem{fu2022asymptotic}
Jinxue Fu, Weiming Li, Peng Song, and Yanli Wang.
\newblock An asymptotic-preserving imex method for nonlinear radiative transfer
  equation.
\newblock {\em Journal of Scientific Computing}, 92(1):27, 2022.

\bibitem{gao2021phygeonet}
Han Gao, Luning Sun, and Jian-Xun Wang.
\newblock Phygeonet: Physics-informed geometry-adaptive convolutional neural
  networks for solving parameterized steady-state pdes on irregular domain.
\newblock {\em Journal of Computational Physics}, 428:110079, 2021.

\bibitem{GENTILE2001543}
N.~A. Gentile.
\newblock Implicit monte carlo diffusion—an acceleration method for monte
  carlo time-dependent radiative transfer simulations.
\newblock {\em Journal of Computational Physics}, 172(2):543--571, 2001.

\bibitem{guo2023pre}
Jiawei Guo, Yanzhong Yao, Han Wang, and Tongxiang Gu.
\newblock Pre-training strategy for solving evolution equations based on
  physics-informed neural networks.
\newblock {\em Journal of Computational Physics}, 489:112258, 2023.

\bibitem{he2016deep}
Kaiming He, Xiangyu Zhang, Shaoqing Ren, and Jian Sun.
\newblock Deep residual learning for image recognition.
\newblock In {\em Proceedings of the IEEE conference on computer vision and
  pattern recognition}, pages 770--778, 2016.

\bibitem{hendrycks2016gaussian}
Dan Hendrycks and Kevin Gimpel.
\newblock Gaussian error linear units (gelus).
\newblock {\em arXiv preprint arXiv:1606.08415}, 2016.
\newblock Published online June 5. Accessed July 13, 2024.

\bibitem{howell2020thermal}
John~R Howell, M~Pinar Meng{\"u}{\c{c}}, Kyle Daun, and Robert Siegel.
\newblock {\em Thermal radiation heat transfer}.
\newblock CRC press, 2020.

\bibitem{jagtap2020extended}
Ameya~D. Jagtap and Georges~Em Karniadakis.
\newblock Extended physics-informed neural networks (xpinns): A generalized
  space-time domain decomposition based deep learning framework for nonlinear
  partial differential equations.
\newblock {\em Communications in Computational Physics}, 28(5):2002--2041,
  2020.

\bibitem{jin2023asymptotic}
Shi Jin, Zheng Ma, and Keke Wu.
\newblock Asymptotic-preserving neural networks for multiscale time-dependent
  linear transport equations.
\newblock {\em Journal of Scientific Computing}, 94(3):57, 2023.

\bibitem{jin2023asymptoticpreserving}
Shi Jin, Zheng Ma, and Keke Wu.
\newblock Asymptotic-preserving neural networks for multiscale kinetic
  equations.
\newblock {\em Communications in Computational Physics}, 35(3):693--723, 2024.

\bibitem{jin1998diffusive}
Shi Jin, Lorenzo Pareschi, and Giuseppe Toscani.
\newblock Diffusive relaxation schemes for multiscale discrete-velocity kinetic
  equations.
\newblock {\em SIAM Journal on Numerical Analysis}, 35(6):2405--2439, 1998.

\bibitem{jin2000uniformly}
Shi Jin, Lorenzo Pareschi, and Giuseppe Toscani.
\newblock Uniformly accurate diffusive relaxation schemes for multiscale
  transport equations.
\newblock {\em SIAM Journal on Numerical Analysis}, 38(3):913--936, 2000.

\bibitem{kingma2014adam}
Diederik~P Kingma.
\newblock Adam: A method for stochastic optimization.
\newblock {\em arXiv preprint arXiv:1412.6980}, 2014.

\bibitem{lathrop1964discrete}
Kaye~D Lathrop and Bengt~G Carlson.
\newblock Discrete ordinates angular quadrature of the neutron transport
  equation.
\newblock Technical report, Los Alamos National Lab.(LANL), Los Alamos, NM
  (United States), 1964.

\bibitem{Li2022AMA}
Hongyan Li, Song Jiang, Wenjun Sun, Liwei Xu, and Guanyu Zhou.
\newblock A model-data asymptotic-preserving neural network method based on
  micro-macro decomposition for gray radiative transfer equations.
\newblock {\em Communications in Computational Physics}, 35(5):1155--1193,
  2024.

\bibitem{li2024asymptotic}
Ruo Li, Weiming Li, Shengtong Liang, Yuehan Shao, Min Tang, and Yanli Wang.
\newblock An asymptotic-preserving method for the three-temperature radiative
  transfer model.
\newblock {\em arXiv preprint arXiv:2402.19191}, 2024.

\bibitem{li2020unified}
Weiming Li, Chang Liu, Yajun Zhu, Jiwei Zhang, and Kun Xu.
\newblock Unified gas-kinetic wave-particle methods iii: Multiscale photon
  transport.
\newblock {\em Journal of Computational Physics}, 408:109280, 2020.

\bibitem{liu1989limited}
Dong~C Liu and Jorge Nocedal.
\newblock On the limited memory bfgs method for large scale optimization.
\newblock {\em Mathematical programming}, 45(1):503--528, 1989.

\bibitem{lu2022solving}
Yulong Lu, Li~Wang, and Wuzhe Xu.
\newblock Solving multiscale steady radiative transfer equation using neural
  networks with uniform stability.
\newblock {\em Research in the Mathematical Sciences}, 9(3):45, 2022.

\bibitem{MCCLARREN20087561}
Ryan~G. McClarren, Thomas~M. Evans, Robert~B. Lowrie, and Jeffery~D. Densmore.
\newblock Semi-implicit time integration for pn thermal radiative transfer.
\newblock {\em Journal of Computational Physics}, 227(16):7561--7586, 2008.

\bibitem{MCCLARREN20082864}
Ryan~G. McClarren, James~Paul Holloway, and Thomas~A. Brunner.
\newblock On solutions to the pn equations for thermal radiative transfer.
\newblock {\em Journal of Computational Physics}, 227(5):2864--2885, 2008.

\bibitem{mcclarren2009modified}
Ryan~G. McClarren and Todd~J Urbatsch.
\newblock A modified implicit monte carlo method for time-dependent radiative
  transfer with adaptive material coupling.
\newblock {\em Journal of Computational Physics}, 228(16):5669--5686, 2009.

\bibitem{mieussens2013asymptotic}
Luc Mieussens.
\newblock On the asymptotic preserving property of the unified gas kinetic
  scheme for the diffusion limit of linear kinetic models.
\newblock {\em Journal of Computational Physics}, 253:138--156, 2013.

\bibitem{miller1991analysis}
Warren~F Miller~Jr.
\newblock An analysis of the finite differenced, even-parity, discrete
  ordinates equations in slab geometry.
\newblock {\em Nuclear Science and Engineering}, 108(3):247--266, 1991.

\bibitem{MISHRA2021107705}
Siddhartha Mishra and Roberto Molinaro.
\newblock Physics informed neural networks for simulating radiative transfer.
\newblock {\em Journal of Quantitative Spectroscopy and Radiative Transfer},
  270:107705, 2021.

\bibitem{morel1996linear}
Jim~E Morel, Todd~A Wareing, and Kenneth Smith.
\newblock A linear-discontinuous spatial differencing scheme forsnradiative
  transfer calculations.
\newblock {\em Journal of Computational Physics}, 128(2):445--462, 1996.

\bibitem{pomraning2005equations}
Gerald~C Pomraning.
\newblock {\em The equations of radiation hydrodynamics}.
\newblock Courier Corporation, 2005.

\bibitem{raissi2019physics}
Maziar Raissi, Paris Perdikaris, and George~E Karniadakis.
\newblock Physics-informed neural networks: A deep learning framework for
  solving forward and inverse problems involving nonlinear partial differential
  equations.
\newblock {\em Journal of Computational physics}, 378:686--707, 2019.

\bibitem{schafer2011diffusive}
Matthias Sch{\"a}fer, Martin Frank, and C~David Levermore.
\newblock Diffusive corrections to p\_n approximations.
\newblock {\em Multiscale Modeling \& Simulation}, 9(1):1--28, 2011.

\bibitem{seibold2014starmap}
Benjamin Seibold and Martin Frank.
\newblock Starmap---a second order staggered grid method for spherical
  harmonics moment equations of radiative transfer.
\newblock {\em ACM Transactions on Mathematical Software (TOMS)}, 41(1):1--28,
  2014.

\bibitem{shi2020continuous}
Yi~Shi, Xiaole Han, Wenjun Sun, and Peng Song.
\newblock A continuous source tilting scheme for radiative transfer equations
  in implicit monte carlo.
\newblock {\em Journal of Computational and Theoretical Transport},
  50(1):1--26, 2021.

\bibitem{shi2023efficient}
Yi~Shi, Peng Song, and Tao Xiong.
\newblock An efficient asymptotic preserving monte carlo method for radiative
  transfer equations.
\newblock {\em Journal of Computational Physics}, 493:112483, 2023.

\bibitem{smedley2015asymptotic}
Richard~P Smedley-Stevenson and Ryan~G McClarren.
\newblock Asymptotic diffusion limit of cell temperature discretisation schemes
  for thermal radiation transport.
\newblock {\em Journal of Computational Physics}, 286:214--235, 2015.

\bibitem{steinberg2022multi}
Elad Steinberg and Shay~I Heizler.
\newblock Multi-frequency implicit semi-analog monte-carlo (ismc) radiative
  transfer solver in two-dimensions (without teleportation).
\newblock {\em Journal of Computational Physics}, 450:110806, 2022.

\bibitem{sun2015asymptotic}
Wenjun Sun, Song Jiang, and Kun Xu.
\newblock An asymptotic preserving unified gas kinetic scheme for gray
  radiative transfer equations.
\newblock {\em Journal of Computational Physics}, 285:265--279, 2015.

\bibitem{SUN2017455}
Wenjun Sun, Song Jiang, and Kun Xu.
\newblock A multidimensional unified gas-kinetic scheme for radiative transfer
  equations on unstructured mesh.
\newblock {\em Journal of Computational Physics}, 351:455--472, 2017.

\bibitem{thomas2002radiative}
Gary~E Thomas and Knut Stamnes.
\newblock {\em Radiative transfer in the atmosphere and ocean}.
\newblock Cambridge University Press, 2002.

\bibitem{walters1991investigation}
WF~Walters and JE~Morel.
\newblock Investigation of linear-discontinuous angular differencing for the
  1-d spherical-geometry s sub n equations.
\newblock Technical report, Los Alamos National Lab., NM (USA), 1991.

\bibitem{wang2021eigenvector}
Sifan Wang, Hanwen Wang, and Paris Perdikaris.
\newblock On the eigenvector bias of fourier feature networks: From regression
  to solving multi-scale pdes with physics-informed neural networks.
\newblock {\em Computer Methods in Applied Mechanics and Engineering},
  384:113938, 2021.

\bibitem{xiong2022high}
Tao Xiong, Wenjun Sun, Yi~Shi, and Peng Song.
\newblock High order asymptotic preserving discontinuous galerkin methods for
  gray radiative transfer equations.
\newblock {\em Journal of Computational Physics}, 463:111308, 2022.

\bibitem{xu2010unified}
Kun Xu and Juan-Chen Huang.
\newblock A unified gas-kinetic scheme for continuum and rarefied flows.
\newblock {\em Journal of Computational Physics}, 229(20):7747--7764, 2010.

\bibitem{XU20171}
Longfei Xu, Liangzhi Cao, Youqi Zheng, and Hongchun Wu.
\newblock Development of a new parallel sn code for neutron-photon transport
  calculation in 3-d cylindrical geometry.
\newblock {\em Progress in Nuclear Energy}, 94:1--21, 2017.

\bibitem{xu2020asymptotic}
Xiaojing Xu, Wenjun Sun, and Song Jiang.
\newblock An asymptotic preserving angular finite element based unified gas
  kinetic scheme for gray radiative transfer equations.
\newblock {\em Journal of Quantitative Spectroscopy and Radiative Transfer},
  243:106808, 2020.

\bibitem{yang2020physics}
Liu Yang, Dongkun Zhang, and George~Em Karniadakis.
\newblock Physics-informed generative adversarial networks for stochastic
  differential equations.
\newblock {\em SIAM Journal on Scientific Computing}, 42(1):A292--A317, 2020.

\bibitem{yu2023mcmc}
Tengchao Yu, Heng Yong, Li~Liu, Han wang, and Hua chen.
\newblock Mcmc-pinns: A modified markov chain monte-carlo method for sampling
  collocation points of pinns adaptively.
\newblock {\em TechRxiv}, February 01, 2023.

\end{thebibliography}

\end{document}